\documentclass[epj,nopacs]{svjour}
\pdfoutput=1
\usepackage{amsmath}
\usepackage{graphicx}
\usepackage{cite}

\newcommand{\nn}{\nonumber}
\newcommand{\sig}{\sigma}
\newcommand{\lam}{\lambda}
\newcommand{\tm}{\tau^-}
\newcommand{\tp}{\tau^+}
\newcommand{\M}{{\cal M}}
\newcommand{\p}{{\cal P}}
\newcommand{\D}{{\cal D}}
\newcommand{\stau}{\tilde\tau_1}
\newcommand{\neu}{\tilde\chi^0_1}

\begin{document}

\title{TauDecay: a library to simulate polarized tau decays\\
       via FeynRules and MadGraph5}

\author{
 Kaoru Hagiwara\inst{1}\fnmsep
 \and
 Tong Li\inst{2}\fnmsep\thanks{e-mail: tong.li@monash.edu}
 \and
 Kentarou Mawatari\inst{3}\fnmsep\thanks{e-mail:
            kentarou.mawatari@vub.ac.be}
 \and
 Junya Nakamura\inst{1}\fnmsep\thanks{e-mail: junnaka@post.kek.jp}
 }

\institute{
 KEK Theory Center, and Sokendai, Tsukuba 305-0801, Japan
\and
ARC Centre of Excellence for Particle Physics at the Terascale,
 School of Physics, Monash University, \\
 Melbourne, Victoria 3800, Australia
\and
 Theoretische Natuurkunde and IIHE/ELEM, Vrije Universiteit Brussel,\\
 and International Solvay Institutes,
 Pleinlaan 2, B-1050 Brussels, Belgium
 }

\abstract{
{\sc TauDecay} is a library of helicity amplitudes to simulate polarized
tau decays, constructed in the {\sc FeynRules} and {\sc MadGraph5}
framework. 
Together with the leptonic mode, the decay library includes the main
hadronic modes, $\tau\to\nu_{\tau}+\pi$, $2\pi$, and $3\pi$, which are
introduced as effective vertices by using {\sc FeynRules}. 
The model file allows us to simulate tau decays when the on-shell tau
production is kinematically forbidden. 
We also demonstrate that all possible correlations among the decay
products of pair-produced taus through a $Z$ boson and a
scalar/pseudoscalar Higgs boson are produced automatically.
The program has been tested carefully by making use of the standard tau
decay library {\sc Tauola}. 
}

\date{}

\titlerunning{{\sc TauDecay}}
\authorrunning{K.~Hagiwara, T.~Li, K.~Mawatari and J.~Nakamura}

\maketitle

\vspace*{-110mm}
\noindent KEK-TH-1596 \\[1mm]
\vspace*{87mm}



\section{Introduction}\label{intro}

The recent observation of a standard-model-like Higgs boson with mass
around 125~GeV at the LHC~\cite{:2012gk} has increased our interests in 
the $H\to\tau^+\tau^-$ decay mode, which has the modest branching
fraction yet to be established~\cite{Chatrchyan:2012vp}.
On the other hand, tau leptons have been well known as a particle which
can explore the Higgs sector~\cite{Bullock:1991fd,Kramer:1993jn} as well
as supersymmetric models~\cite{Nojiri:1994it,Choi:2006mt}, through their
polarization~\cite{Tsai:1971vv,Hagiwara:1989fn,Rouge:1990kv}.

{\sc Tauola}~\cite{Jadach:1990mz,Jezabek:1991qp,Jadach:1993hs} is a
well-known program to simulate polarized $\tau$ decays, including
various hadronic modes, and recent theoretical and experimental event
simulations involving $\tau$ decays heavily depend on it.%
\footnote{The event generators, {\sc Herwig++} and {\sc Pythia8}, have
their own $\tau$-decay package~\cite{Grellscheid:2007tt,Ilten:2012zb};
{\sc TauSpinner} for studies on spin effect in tau production was
recently reported~\cite{Czyczula:2012ny}.}
The standard {\sc Tauola Universal Interface}~\cite{Golonka:2003xt}
includes the longitudinal spin correlations between pair-produced
taus~\cite{Pierzchala:2001gc}, while the extended version also takes
into account the transverse spin
effects~\cite{Was:2002gv,Bower:2002zx}. 
We point here some limitations of the use of {\sc Tauola}:
\begin{enumerate}
 \item[i)] The produced $\tau$ has to be stable, i.e. on-shell, in event
	 generators.  
 \item[ii)] The interface for the spin correlation effects is limited to 
	  the standard processes such as $Z/\gamma^*\to\tau^+\tau^-$ and
	  $H/A\to\tau^+\tau^-$.
 \item[iii)] For the transverse spin effects, a particular simulation is 
	   required by accepting or rejecting a pair of produced taus
	   based on a spin weighting factor.
\end{enumerate}

In this article, we present a new implementation of a $\tau$-decay model 
file, and construct a library of helicity amplitudes, {\sc TauDecay},%
\footnote{The {\sc TauDecay} package is supported by 
 {\sc MadGraph5\_ aMC@NLO}~\cite{Alwall:2014hca}; see more details in 
`Note added'.}
to simulate polarized $\tau$ decays in the {\sc FeynRules}
(FR)~\cite{Christensen:2008py} and {\sc MadGraph5}
(MG5)~\cite{Alwall:2011uj} framework.
In Section~\ref{sec:fr} we describe how we implement the effective
vertices via {\sc FR} for the main hadronic decay modes, namely
$\tau\to\nu_{\tau}+\pi$, $2\pi$, and $3\pi$, and
present the validation of our {\sc TauDecay} package.
To address the limitations mentioned above, in Sect.~\ref{sec:stau} we
study scalar-tau (stau) decays in scenarios with stau being nearly
degenerate in mass with neutralino, where the on-shell tau production is
kinematically forbidden. 
Moreover, in Sections~\ref{sec:spin} and \ref{sec:taudecay} we
demonstrate that all possible correlations among the decay products of
pair-produced taus via a $Z$ boson and a scalar/pseudoscalar Higgs boson
can be produced within our full-fledged package.
Sect.~\ref{sec:summary} presents our brief summary.

\section{Effective vertices for hadronic tau decays}\label{sec:fr}

In this section, we consider effective vertices for the three hadronic
$\tau$ decay modes, which together with the leptonic mode account for
90\% of the $\tau$ decays~\cite{Beringer:1900zz}, and describe how we
implemented them via the {\sc FeynRules} 
({\sc FR})~\cite{Christensen:2008py} and {\sc MadGraph5} 
({\sc MG5})~\cite{Alwall:2011uj} packages.

\begin{table}
\center
\begin{tabular}{cll}\hline
 $G_F$ & Fermi constant &$1.166379\times10^{-5}$ [GeV$^{-2}$] \\
 $f_{\pi}$ & pion decay constant & 0.13041 [GeV] \\
 $\cos\theta_C$ & Cabibbo angle & 0.97418 \\\hline
\end{tabular}
\caption{Physical constants~\cite{Beringer:1900zz}.}
\label{tab:const}
\end{table}

The decay modes which we consider in this article are
\begin{subequations}
\begin{align}
 &\pi\ {\rm mode}:  &&\tm\to\nu_{\tau}\pi^-, \\
 &\rho\ {\rm mode}: &&\tm\to\nu_{\tau}\rho^-
  \to\nu_{\tau}\pi^-\pi^0, \\
 &a_1\ {\rm mode}:  &&\tm\to\nu_{\tau}a_1^-
  \to\nu_{\tau}\pi^0\rho^-\to\nu_{\tau}\pi^0\pi^0\pi^-, \\
  &                 &&\tm\to\nu_{\tau}a_1^-
  \to\nu_{\tau}\pi^-\rho^0\to\nu_{\tau}\pi^-\pi^-\pi^+,
\end{align}
\end{subequations}
as the 2$\pi$ and 3$\pi$ modes are dominated by the $\rho$ and $a_1$ 
vector-meson productions, respectively.

The $\tau$-$\nu_{\tau}$-$\pi$ vertex can be introduced by the effective
interaction Lagrangian:
\begin{align}
 {\cal L}_{\pi}&= \sqrt{2}G_F f_1\,
  \bar{\tau}\gamma^{\mu} P_L \nu_{\tau}\,\partial_{\mu}\pi^-+h.c.
\label{L_pi}
\end{align}
with the physical constants in Table~\ref{tab:const}, the chiral
projection operator $P_L$, and the constant form factor
\begin{align}
 f_1=f_{\pi}\cos\theta_C.
\end{align}

The effective Lagrangian for the $\rho$ mode is
\begin{align}
 {\cal L}_{\rho}&= \sqrt{2}G_Ff_2\,
  \bar{\tau}\gamma^{\mu} P_L \nu_{\tau}\, 
  (\pi^0\partial_{\mu}\pi^- - \pi^-\partial_{\mu}\pi^0)+h.c.
\label{L_rho}
\end{align}
with
\begin{align}
 f_2=\sqrt{2}\cos\theta_CF_{\rho}(Q^2).
\label{f2}
\end{align}
The form factor $F_{\rho}(Q^2)$ is parametrized
by~\cite{Kuhn:1990ad,Jadach:1990mz}
\begin{align}
 F_{\rho}(Q^2)=[B_{\rho}(Q^2)+\alpha B_{\rho'}(Q^2)]/(1+\alpha)
\label{F_rho}
\end{align}
with the Breit-Wigner factor
\begin{align}
 B_{V}(Q^2)=\frac{m_{V}^2}
  {m_{V}^2-Q^2-i\sqrt{Q^2}\Gamma_{V}(Q^2)},
\end{align}
where $Q = q_1 + q_2$ for
$\tau^-\to\nu\pi^-(q_1)\pi^0(q_2)$. The running width is
\begin{align}
 \Gamma_{V}(Q^2)=\Gamma_{V}\frac{\sqrt{Q^2}}{m_{V}}
  \frac{g_V(Q^2)}{g_V(m_{V}^2)},
\label{rwidth}
\end{align}
where the $\rho$ meson line shape factor is
\begin{align}
g_{\rho}(Q^2)=\bar\beta
 \Big(\frac{m_{\pi^-}^2}{Q^2},\frac{m_{\pi^0}^2}{Q^2}\Big)^3
\end{align}
with
\begin{align}
 \bar\beta(a,b)\equiv(1+a^2+b^2-2a-2b-2ab)^{1/2}.
\label{beta}
\end{align}
We take $\alpha=-0.145$ in \eqref{F_rho}~\cite{Kuhn:1990ad}.

In practice, we introduce only pions, $\pi^{\pm}$ and $\pi^0$, as new 
particles and implement the above Lagrangians \eqref{L_pi} and
\eqref{L_rho} into {\sc FR},%
\footnote{$\pi^0$ decay is not considered in the program.}
which provides the {\sc Ufo} ({\sc Universal FeynRules Output}) model
file~\cite{Degrande:2011ua}, with $f_2$ as a constant parameter.
The {\sc Aloha} ({\sc Automatic Libraries Of Helicity Amplitudes})
program~\cite{deAquino:2011ub} in {\sc MG5} reads the model file to
create the {\sc Helas} ({\sc HELicity Amplitude Subroutines})
library~\cite{Hagiwara:1990dw} for helicity amplitude computations. 
At this stage we replace the constant parameter by the
momentum-dependent form factor~\eqref{f2}. 

Unlike the above two cases, the effective vertex for the $a_1$ mode
cannot be obtained by the Lagrangian since the vertex structure is not 
symmetric between the two identical pion in the final state.
Therefore, instead of introducing the Lagrangian, we implement it
by hand in {\sc FR} and created the {\sc Ufo} model file.%
\footnote{{\tt Input} option of the function {\tt WriteUFO} in {\sc FR} 
 allows us to include such a vertex~\cite{Degrande:2011ua}.}
The effective vertex, or the decay amplitude, for the $a_1$ mode is
\begin{align}
 {\cal M}_{a_1}=\sqrt{2}G_F\,
 \bar{\tau}\gamma^{\mu}P_L\nu_{\tau}J_{\mu},
\end{align}
where the hadronic current $J^{\mu}$ is given
by~\cite{Kuhn:1990ad,Jadach:1990mz}
\begin{align}
 J^{\mu}=f_3\big[F^{13}(q_1^{\mu}-q_3^{\mu}-G^{13}Q^{\mu})
 + (1\leftrightarrow2)\big]
\end{align}
with $Q=q_1+q_2+q_3$ for
$\tau^-\to\nu_{\tau}\pi^-(q_1)\pi^-(q_2)\pi^+(q_3)$.
The form factors are
\begin{align}
 &f_3=\frac{4}{3f_{\pi}}\cos\theta_CB_{a_1}(Q^2), \\
 &F^{i3}=F_{\rho}(Q_{i3}^2),\quad
 G^{i3}=\frac{Q\cdot(q_i-q_3)}{Q^2},
\end{align}
with $Q_{i3}=q_i+q_3\ (i=1,2)$. The $a_1$ meson line shape factor
in~\eqref{rwidth} is parametrized as~\cite{Kuhn:1990ad,Jadach:1990mz} 
\begin{align}
 g_{a_1}(Q^2)=
 \begin{cases}
  \dfrac{4.1}{Q^2}(Q^2-9m_{\pi}^2)^3 \\
    \ \times\big[1-3.3(Q^2-9m_{\pi}^2)+5.8(Q^2-9m_{\pi}^2)^2\big] \\
   \hspace*{3.5cm}{\rm if}\ Q^2<(m_{\rho}+m_{\pi})^2, \\
  1.623+\dfrac{10.38}{Q^2}-\dfrac{9.32}{Q^4}+\dfrac{0.65}{Q^6} \\
   \hspace*{3.5cm}{\rm if}\ Q^2>(m_{\rho}+m_{\pi})^2.
 \end{cases}
\end{align}
The $\pi^0\pi^0\pi^-$ mode is the same, except the mass difference
between $\pi^{\pm}$ and $\pi^0$.

\begin{table}
\center
\begin{tabular}{cll}\hline
 particle & mass [GeV] & width [GeV] \\ \hline
 $\tau$ & 1.77682 & 2.265$\times10^{-12}$ \\
 $\pi^{\pm}$ & 0.1395702 \\
 $\pi^0$ & 0.1349766 \\
 $\rho$ & 0.77549 & 0.1491 \\
 $\rho(1450)\equiv\rho'$ & 1.465 & 0.4 \\
 $a_1$ & 1.23 & 0.42 \\ \hline
\end{tabular}
\caption{Particle masses and widths~\cite{Beringer:1900zz}.}
\end{table}

\begin{figure}
\center
 \includegraphics[width=.155\textwidth,clip]{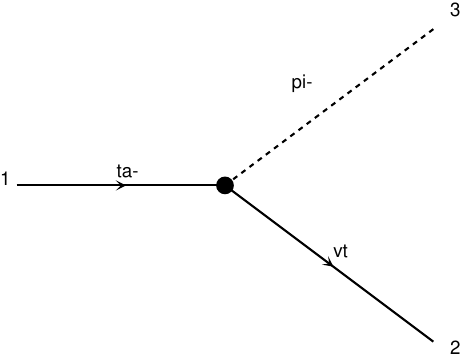}
 \includegraphics[width=.155\textwidth,clip]{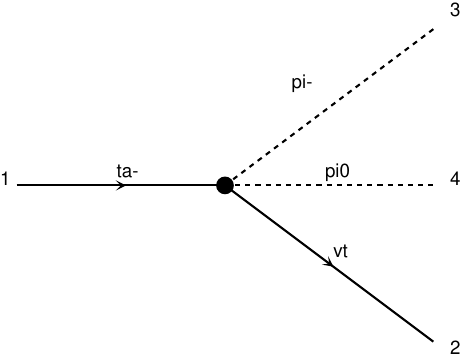}
 \includegraphics[width=.155\textwidth,clip]{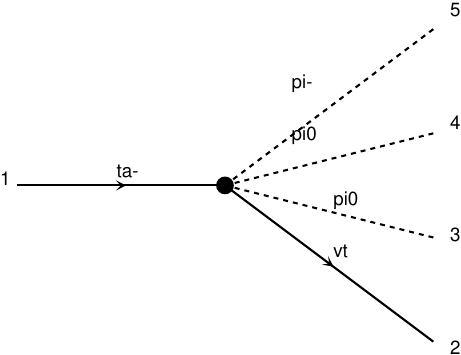}
 \caption{Effective vertices generated by {\sc MG5}.}
\label{fig:vertex}
\end{figure}

\begin{table}
\center
\begin{tabular}{ccccc}\hline
 mode &&  \multicolumn{3}{c}{width [10$^{-13}$~GeV]} \\
  && PDG~\cite{Beringer:1900zz}
  & {\sc Tauola}~\cite{Jadach:1993hs}  & {\sc TauDecay} \\ \hline
 $e^-\bar\nu$ && 4.04 & 4.02 & 4.04 \\
 $\mu^-\bar\nu$ && 3.94 & 3.91 & 3.94 \\
 $\pi^-$  && 2.45 & 2.47 & 2.42 \\
 $\pi^-\pi^0$ && 5.78 & 5.39 & 5.39 \\
 $\pi^0\pi^0\pi^-$ && 2.11 & 2.25 & 2.27 \\
 $\pi^-\pi^-\pi^+$ && 2.04 & 2.21 & 2.22 \\ \hline
\end{tabular}
\caption{$\tau$ decay partial widths.}
\label{tab:width}
\end{table}

\begin{figure}
\center
 \includegraphics[width=.24\textwidth,clip]{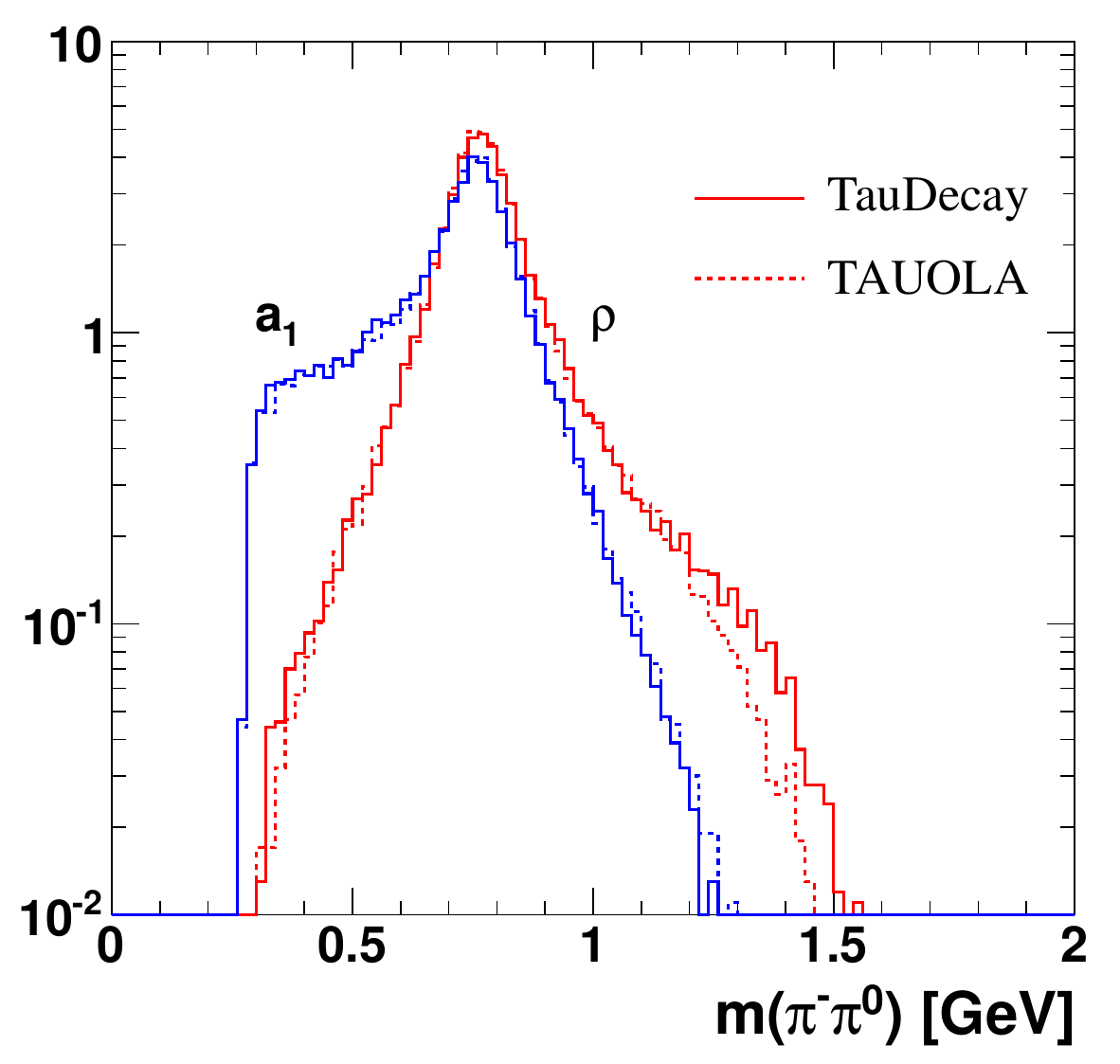}
 \includegraphics[width=.24\textwidth,clip]{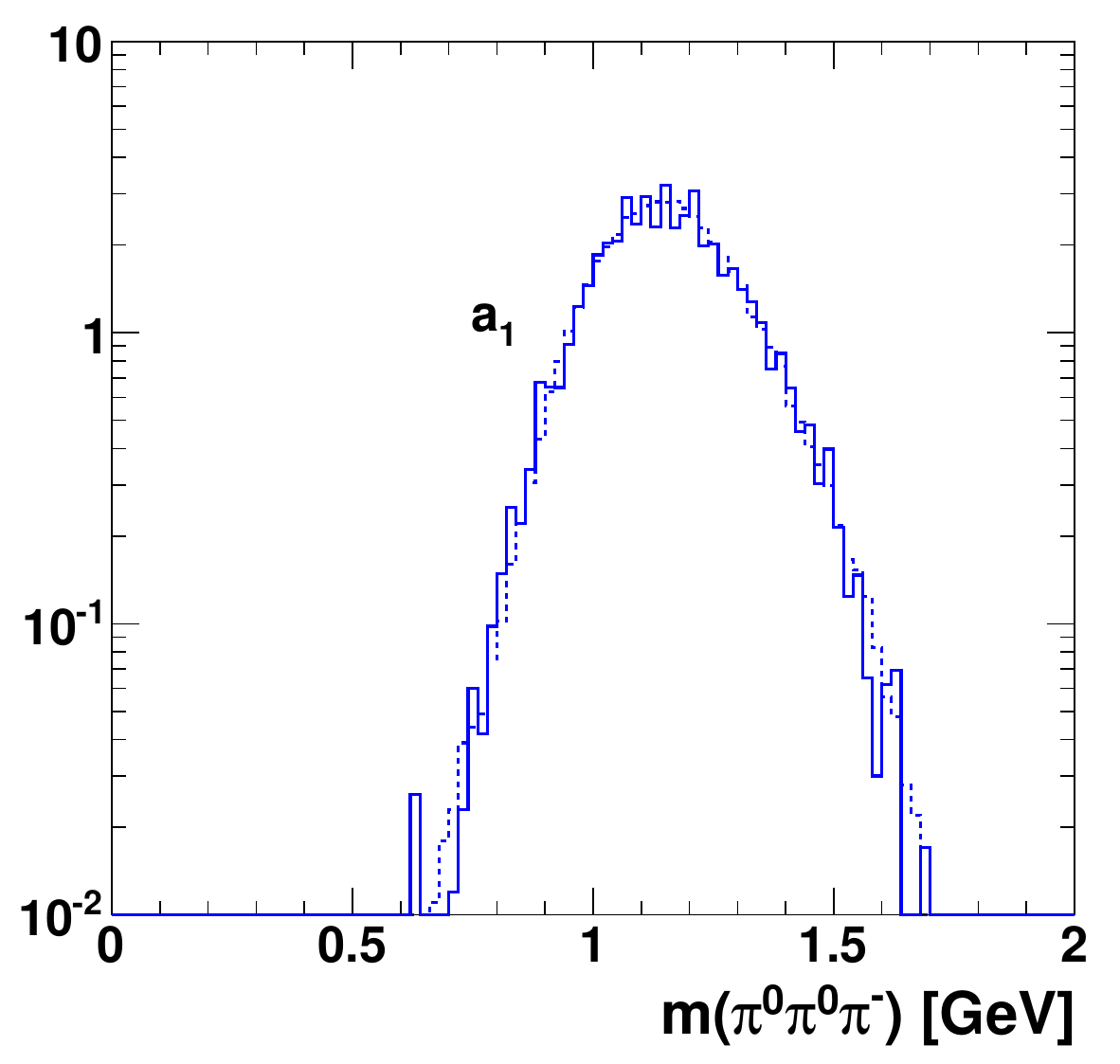}
 \caption{$\pi^-\pi^0$ (left) and $\pi^0\pi^0\pi^-$ (right) invariant 
 mass distributions for the $\tau$ decay into two and three pions.}
\label{fig:minv}
\end{figure}

\begin{figure}
\center
 \includegraphics[width=.24\textwidth,clip]{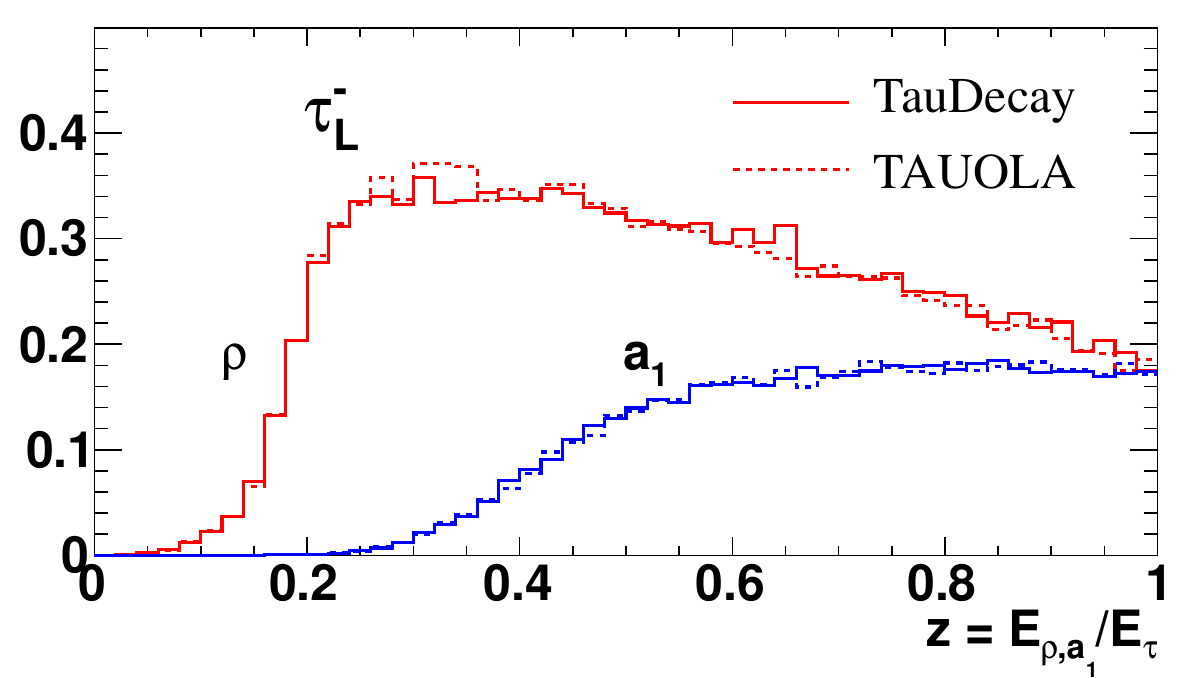}
 \includegraphics[width=.24\textwidth,clip]{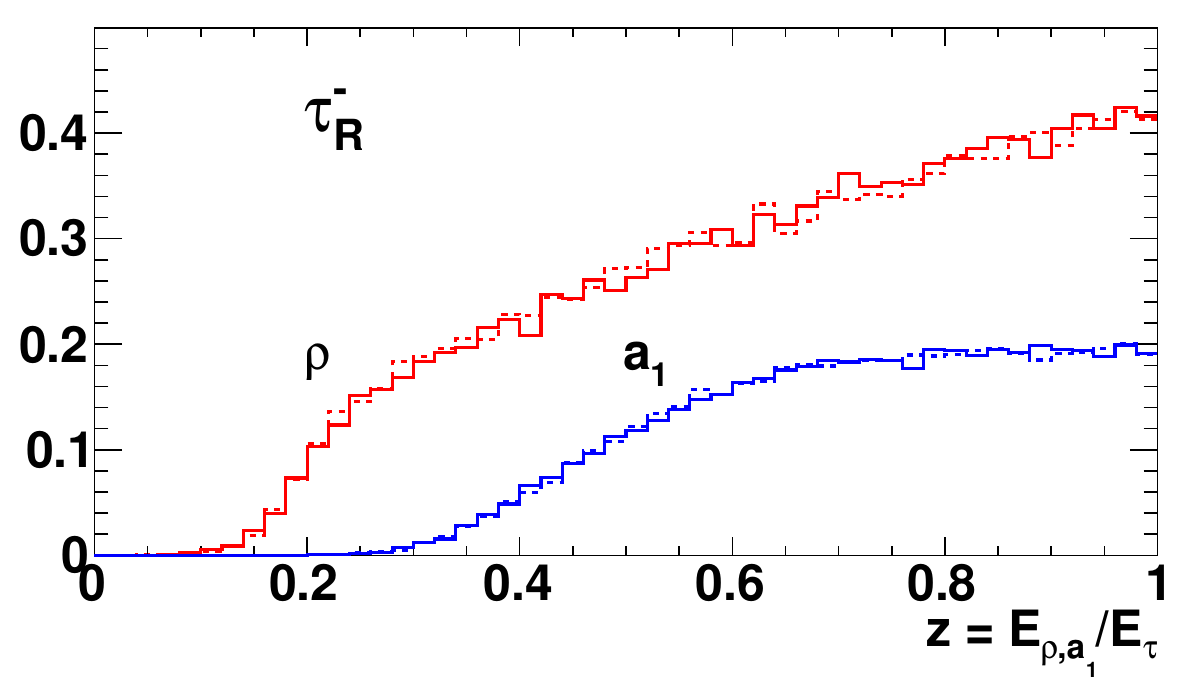}
 \includegraphics[width=.24\textwidth,clip]{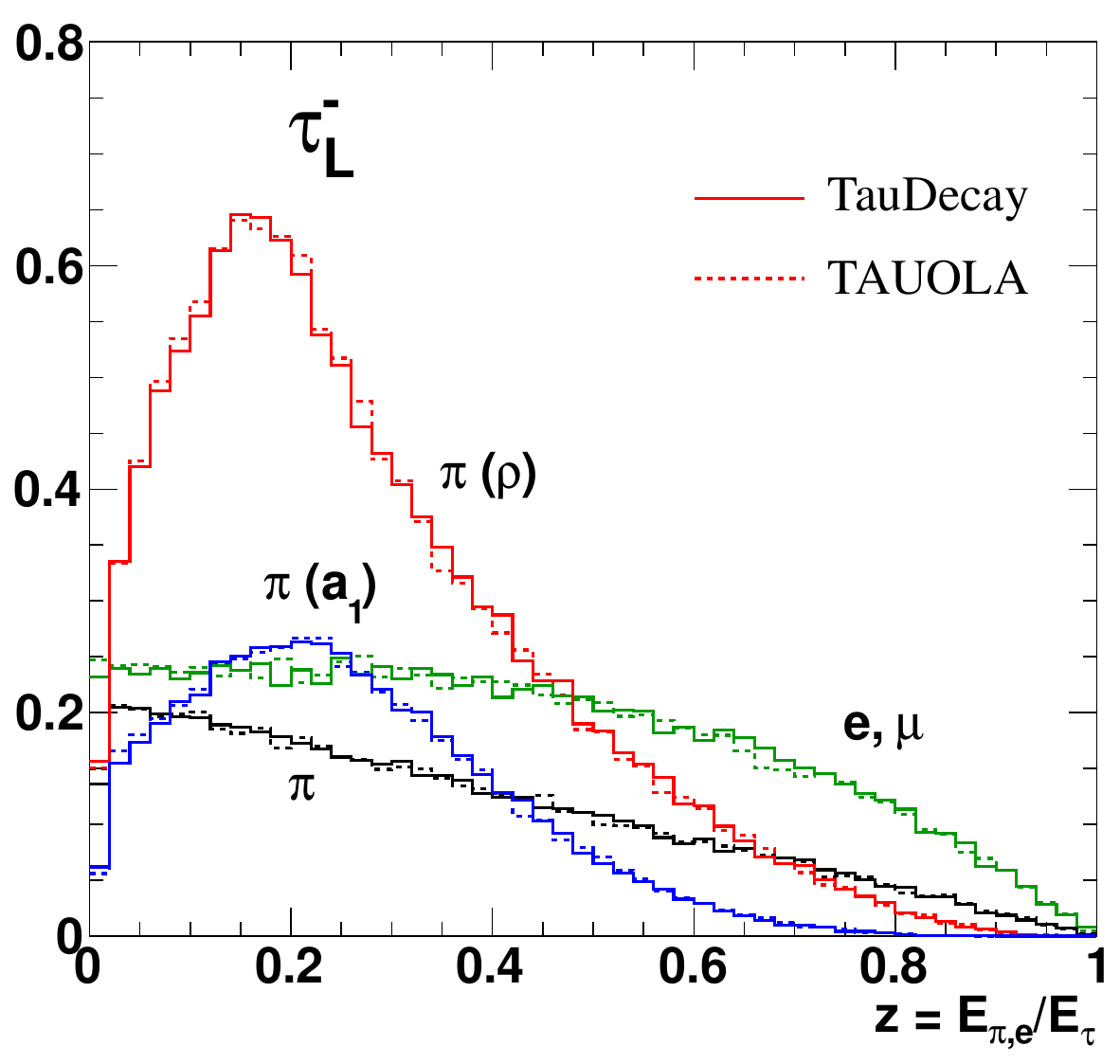}
 \includegraphics[width=.24\textwidth,clip]{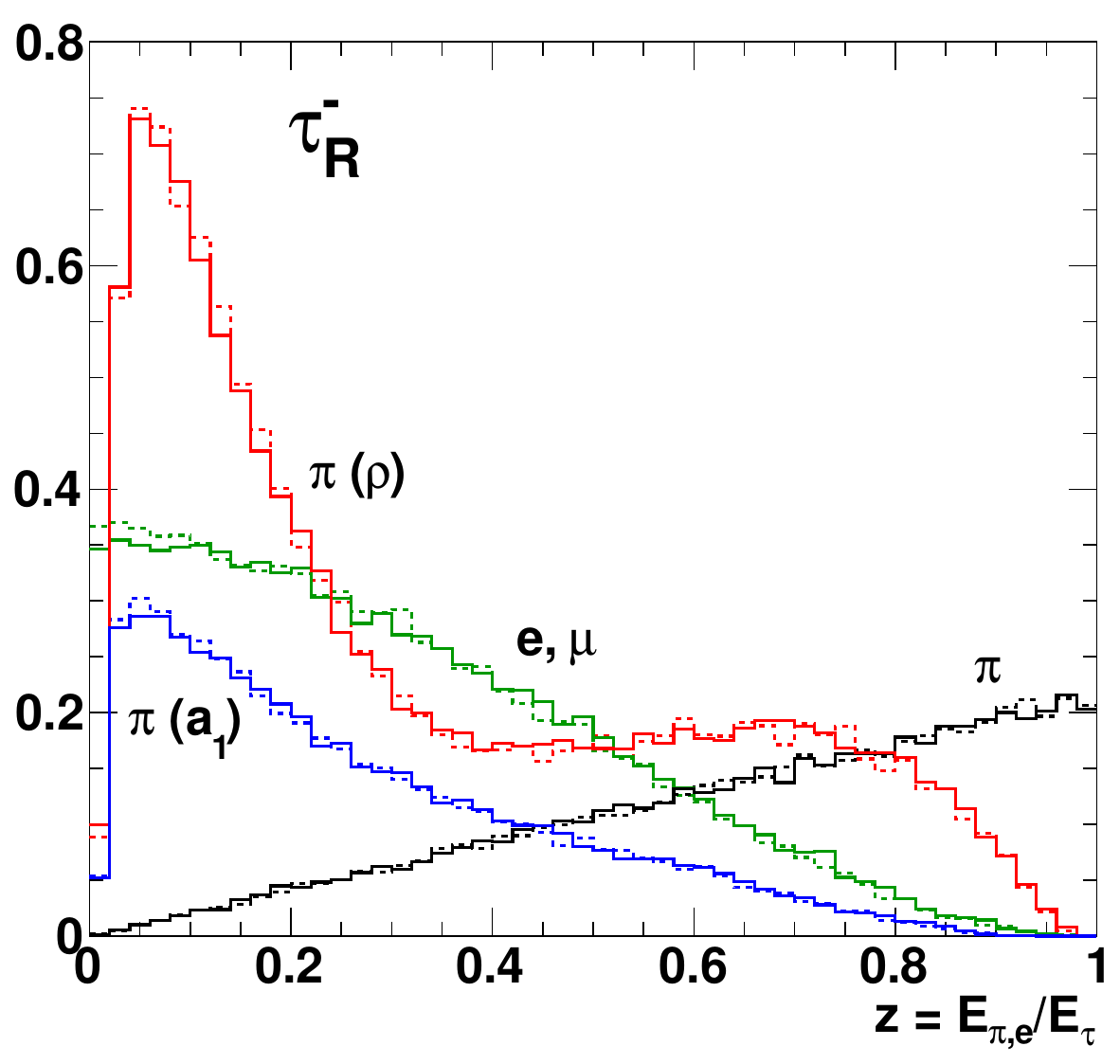}
 \caption{The fractional energy distributions of $\rho$ and $a_1$ (top) 
 and 1-prong $\pi^-$ (bottom) from left-handed $\tau^-$ (left) and
 right-handed $\tau^-$ (right) at $E_{\tau}=50$~GeV, normalized to the
 respective branching ratios. The leptonic decay mode is also shown. 
 The energy fractions are $z\equiv E_{\rho,a_1}/E_{\tau}$ (top) and 
 $E_{\pi,e}/E_{\tau}$ (bottom) in the laboratory frame.}
\label{fig:z}
\end{figure}

Our {\sc Ufo} model now allows {\sc MG5} to generate Feynman diagrams
including hadronic $\tau$ decays (Fig.~\ref{fig:vertex}) and the
corresponding helicity amplitudes, while the leptonic decay mode can be
simulated within the default Standard Model {\sc Ufo} via the off-shell
$W$ boson. 
{\sc TauDecay} is a library to
simulate polarized $\tau$ decays, based on the $\tau$ decay helicity
amplitudes created by {\sc MG5} with the form factor implementation.
Using {\sc TauDecay}, we produce the partial decay widths in
Table~\ref{tab:width} as well as the pion invariant mass distributions
in Fig.~\ref{fig:minv} in good agreement with 
the standard $\tau$ decay
library {\sc Tauola}~\cite{Jadach:1993hs}.%
\footnote{The hadronic currents in {\sc TauDecay} are same as in 
{\sc Tauola}~\cite{Jadach:1993hs}, while the
new hadronic currents based on the resonance chiral
Lagrangian have been recently implemented in 
{\sc Tauola}~\cite{Shekhovtsova:2012ra}.}
We note that the slight mismatch between the two programs in
the table as well as in the large $m_{\pi^-\pi^0}$ invariant mass region
for the $\rho$ mode comes from the QED correction for the leptonic mode
in {\sc Tauola} and the different parameter choice, e.g. $f_{\pi}$ and
$m_{\rho'}$. 
Moreover, the
fractional energy distributions of the polarized $\tau$ decays in
Fig.~\ref{fig:z} agree with the ones in the collinear
limit~\cite{Bullock:1991fd} as well as by {\sc Tauola}.  

\section{Stau decays via an off-shell tau}\label{sec:stau}

An important application of our {\sc FR} $\tau$-decay model file is for
the case that on-shell $\tau$ production is kinematically forbidden. The
model file allows {\sc MG5} to treat an intermediate $\tau$ as a
propagator. 

Such a case can happen in the so-called stau-neutralino ($\stau$-$\neu$) 
coannihilation scenario in the constrained minimal supersymmetric
standard model (CMSSM). The neutralino $\neu$ is the lightest
supersymmetric particle (LSP) and stable by assuming the $R$-parity
conservation. The scalar-tau (stau) $\stau$ is the next-to-lightest SUSY
particle (NLSP), which exclusively decays into a $\tau$-lepton and a LSP
$\neu$. The scenario is cosmologically preferred to provide the observed
dark matter relic density~\cite{Ellis:2002rp} as well as to solve the
${}^7$Li problem in the standard big-bang
nucleosynthesis~\cite{Jittoh:2008eq}. The recent global fit analysis for
the CMSSM also favours the point~\cite{Citron:2012fg}. If
$m_{\stau}-m_{\neu}>m_{\tau}$, there is no obstacle to use {\sc Tauola}
for $\tau$ decays, following event generation with $\tau$-leptons as an
on-shell particle~\cite{Arnowitt:2008bz}. For the case of
\begin{align}
 \Delta m=m_{\stau}-m_{\neu}<m_{\tau},
\end{align}
however, we cannot generate events unless $\tau$ decays are taken into
account in event generators. One of the common benchmark points for the
SUSY searches at the LHC is e.g. CMSSM40.1.2~\cite{AbdusSalam:2011fc},
which provides the above mass spectrum with $m_{\stau}=230.3$~GeV and
$m_{\neu}=228.7$~GeV~\cite{Allanach:2001kg}.%
\footnote{A similar situation in the chargino-neutralino degenerate
 scenario has been studied in~\cite{Chen:1996ap,Barr:2002ex}.}

\begin{figure}
\center
 \includegraphics[width=.3\textwidth,clip]{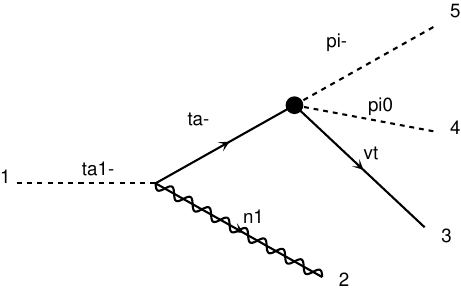}
 \caption{Feynman diagram of a four-body $\stau$ decay generated by 
  {\sc MG5}.} 
\label{fig:stau}
\end{figure}

\begin{figure}
\center
\includegraphics[width=.45\textwidth,clip]{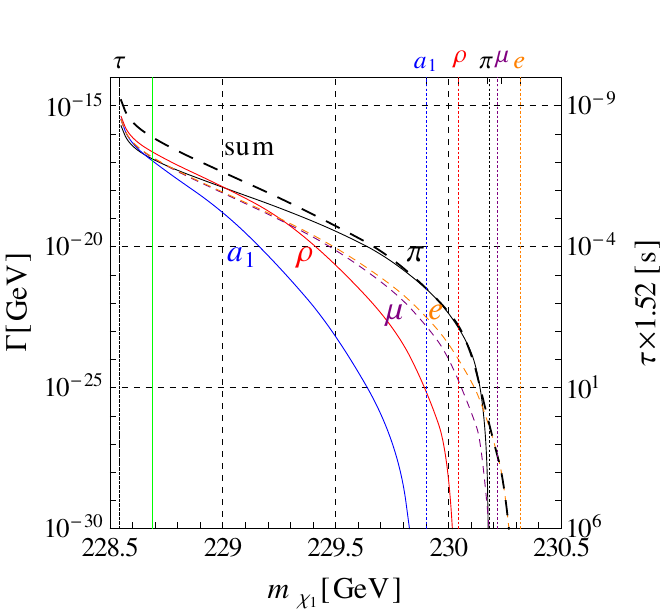}
\caption{Stau partial decay widths as a function of the LSP neutralino 
 mass. The sum of the partial widths is also shown by a black dashed
 line. The stau mass is fixed at 230.3~GeV, based on the CMSSM40.1.2
 benchmark point. The vertical dashed lines denote the threshold for
 each decay mode.} 
\label{fig:stauwidth}
\end{figure}

When the two-body decay mode $\stau\to\neu+\tau$ is closed, the
three-body or four-body decays via the
off-shell $\tau$ (see, e.g. Fig.~\ref{fig:stau}) become dominant and
the $\stau$ could be long-lived~\cite{Profumo:2004qt,Jittoh:2005pq,Citron:2012fg}.
Figure~\ref{fig:stauwidth} shows the stau partial decay widths for each
$\tau$ decay mode
against the $\tilde{\chi}^0_1$ mass. The $\stau$ mass
and the relevant mixing angles in the
neutralino and stau sectors are fixed as in CMSSM40.1.2, providing a
$\tilde\tau_R$-like stau and a bino-like neutralino.
The vertical dashed lines denote the threshold for each decay mode,
where $a_1$ and $\rho$ indicate $\Delta m=m_{3\pi}$ and $m_{2\pi}$, respectively.
As the $\neu$ mass is approaching the $\stau$ mass, i.e. $\Delta m$
becomes smaller, the decay width becomes smaller very rapidly.
Just below the $\tau$ threshold, around the CMSSM40.1.2 point (denoted by
a vertical solid line), decay widths of all the modes are comparable with
those in the on-shell $\tau$ case, while
the $\pi$ mode is dominant in the small $\Delta m$
region due to the phase space suppression for the multi-pion and
leptonic modes.
Only in the $\Delta m < m_{\pi}$ region the electronic mode becomes
dominant.
Those agree with the recent study~\cite{Citron:2012fg}, although $\rho$
and $a_1$ are considered as on-shell particles in their calculations.

The collider signature of the long-lived $\stau$ significantly depends
on the $\stau$ width, i.e. the lifetime; see
e.g. Refs.~\cite{Ambrosanio:2000ik,Ishiwata:2008tp,Asai:2009ka}, and as
shown in Fig.~\ref{fig:stauwidth} the lifetime strongly depends on
$\Delta m$. For instance, the $\stau$ predicted at CMSSM40.1.2, whose
lifetime is ${\cal O}(10^{-8}\,{\rm s})$, could leave a charged track
with a displaced vertex of its decay inside the detectors. We briefly
study the decay distributions for such a long-lived stau to see the
left-right mixing of $\stau$ in this situation.

The scalar partners of left-handed and right-handed taus,
$\tilde{\tau}_L$ and $\tilde{\tau}_R$, mix to form two mass eigenstates,
and the lighter one is 
\begin{align}
 \tilde{\tau}_1 = \cos\theta_{\tilde\tau}\,\tilde\tau_L +
                  \sin\theta_{\tilde\tau}\,\tilde\tau_R,
\end{align}
where $\theta_{\tilde\tau}$ is the mixing angle. Similarly gauginos,
$\tilde{B}$ and $\tilde{W}_3$, and neutral Higgsinos, $\tilde{H}_d^0$
and $\tilde{H}_u^0$, mix to form four mass eigenstates, and the lightest 
one is 
\begin{align}
 \tilde{X}_i = U_{i1}\tilde{\chi}^0_1,
\end{align}
where $\tilde{X}=(\tilde{B},\tilde{W}_3,\tilde{H}_d^0,\tilde{H}_u^0)$. 
The chirality of $\tau$ in the $\tilde{\tau}_1\to \tau\neu$ decay
depends on these mixings and determined by the interaction Lagrangian
\begin{align}
 {\cal L} = \overline{\tilde{\chi}_1^0}(a_LP_L+a_RP_R)\tau\,
  \tilde{\tau}_1^* + h.c. 
\label{eq101}
\end{align}
with the chiral-projection operators,
$P_{R/L}=\frac{1}{2}(1\pm\gamma^5)$. The couplings, $a_L$ and $a_R$, are 
\begin{subequations}
\begin{align}
 a_L& = \cos\theta_{\tilde\tau}\frac{g}{\sqrt{2}} 
  (U_{21}+U_{11}\tan\theta_W)
  + \sin\theta_{\tilde\tau}\,U_{31}Y_{\tau}, \label{eq102} \\
 a_R& = -\sin\theta_{\tilde\tau}\frac{g}{\sqrt{2}}2 U_{11}^*\tan\theta_W
  + \cos\theta_{\tilde\tau}\,U_{31}^*Y_{\tau}, \label{eq103}
\end{align}
\end{subequations}
where $Y_{\tau}=-gm_{\tau}/\sqrt{2}m_W\cos\beta$, $g$ is the $SU(2)_L$
gauge coupling, and $\tan\beta$ is the ratio of the vacuum expectation
values of the two Higgs doublets. For most of SUSY scenarios as well as
for CMSSM40.1.2 the lightest neutralino is gaugino-like 
($U_{31}\sim0$), where the chirality is conserved. 

\begin{figure}
\center
 \includegraphics[width=.24\textwidth,clip]{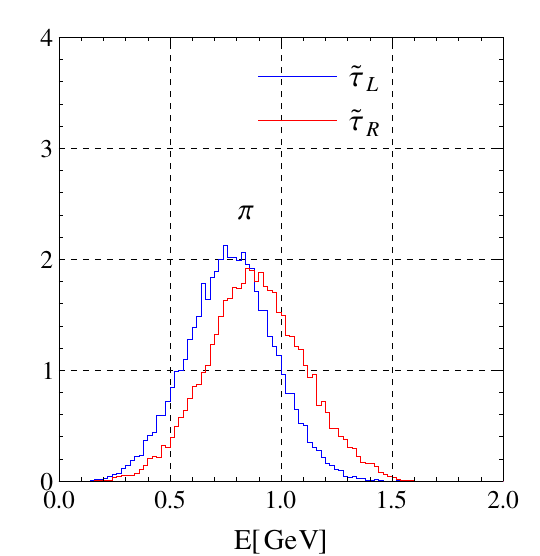}
 \includegraphics[width=.24\textwidth,clip]{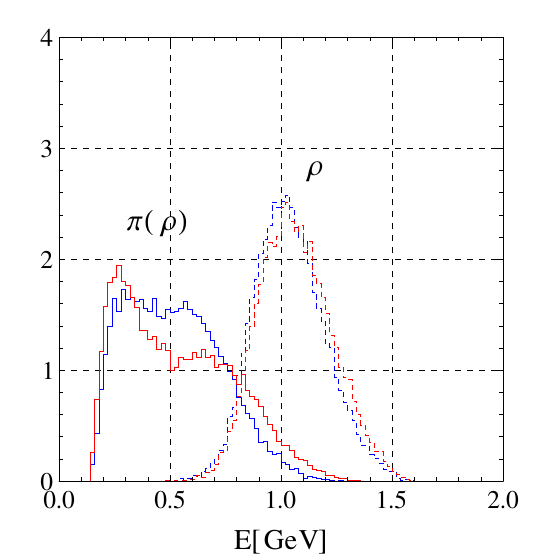}
 \caption{Energy distributions of the one-prong pion in the $\pi$ mode 
 (left) and in the $\rho$ mode (right) for the left-handed
 $\tilde\tau$ (blue) and right-handed $\tilde\tau$ (red) decay in the
 stau rest frame. The distributions for the intermediate $\rho$ is also
 shown by dashed lines as a reference.} 
\label{fig:stau_lr}
\end{figure}

Figure 6 shows the energy distributions of the one-prong pion  in the
$\pi$ mode (left) and in the $\rho$ mode (right) for the left-handed
$\tilde\tau$ (blue) and right-handed $\tilde\tau$ (red) decay in the
stau rest frame. The distributions for the reconstructed $\rho$ is also
shown by dashed lines as a reference. The phase space of the decaying
off-shell $\tau$ is limited and it is no longer energetic. Therefore the
energy distributions of the pion are quite different from the ones in
Fig.~\ref{fig:z}, where the collinear limit is safely applied, although
the remnant of the difference between left-handed and right-handed can
be still seen especially for the $\pi$ and $\rho$ modes.

\section{Spin correlations in tau pair decays}\label{sec:spin}

Another important application of the {\sc FR} $\tau$-decay model is spin
correlations in tau-pair 
production~\cite{Bullock:1991fd,Kramer:1993jn,Pierzchala:2001gc,Was:2002gv,Bower:2002zx,Grellscheid:2007tt,Ilten:2012zb,Czyczula:2012ny}.%
\footnote{The correlation of $\tau^{\pm}\tau^{\pm}$ from a doubly
 charged Higgs boson $H^{\pm\pm}$ has also been studied
 in~\cite{Sugiyama:2012yw}.} 
As in the previous section, one can generate amplitudes including the
$\tau$ decays, and hence the spin correlations between the two taus are
automatically taken into account.
In other words, there is no need for a particular
simulation to get the transverse spin effects. 

In this section, we present correlations in tau-pair production through
a $Z$ boson, a $CP$-even scalar $H$, or a $CP$-odd scalar $A$,
\begin{subequations}
\begin{align}
 q\bar q  &\to Z\to\tau^+\tau^-, \label{processZ}\\
 gg &\to H\to\tau^+\tau^-, \label{processH}\\
 gg &\to A\to\tau^+\tau^-, \label{processA}
\end{align}
\end{subequations}
and the subsequent hadronic $\tau$ decays. We note that, although we
consider the above standard $\tau$-pair production processes in this
paper, a special interface to cooperate with any new physics models is
not required when we generate amplitudes with intermediate taus being as
propagators. 


\subsection{Helicity formalism}

First, we discuss the case that both taus decay to $\nu_{\tau}\pi$ by
using analytic expressions. We assign the four-momentum and the helicity
of each particle for the process as
\begin{align}
 a_1(p_1,&\sig_1)+a_2(p_2,\sig_2) \nn\\
  &\to\tm(q_1,\lam_1)+\tp(q_2,\lam_2) \nn\\
  &\to\pi^-(k_1)+\pi^+(k_2)+\nu_{\tau}(k_3,-)+\bar\nu_{\tau}(k_4,+),
\end{align}
where $a_{1,2}$ stand for quarks or gluons (or even electrons or photons
in case of $e^+e^-$ or $\gamma\gamma$ collisions). The helicities take
the values $\sig_i/2$ and $\lam_i/2$ for quarks and leptons and $\sig_i$
for gluons and photons with $\sig_i=\pm 1$ and $\lam_i=\pm 1$. Although
we evaluate the full amplitude for the above processes numerically to
present the distributions, we discuss it below in terms of the
$\tau$-pair production amplitude and the $\tau$ decay amplitudes, which
give us better understanding of the distributions. Moreover, we
construct our {\sc TauDecay} package based on the following analytic
expressions in the end of this section. 

Using the completeness relations
\begin{subequations}
\begin{align}
 \slash\!\!\!q_1+m&=\sum_{\lam_1}u(q_1,\lam_1)\bar u(q_1,\lam_1), \\
 \slash\!\!\!q_2-m&=\sum_{\lam_2}v(q_2,\lam_2)\bar v(q_2,\lam_2),
\end{align}
\end{subequations}
the full amplitude can be expressed as the product of the tau-pair
production amplitude ($\M_{X=Z,H,A}$) and two $\tau\to\pi\nu$ decay
amplitudes ($\M_{1,2}$):
\begin{align}
 &\M(p_1,\sig_1;p_2,\sig_2;k_i) \nn\\
 &=D(q_1^2)D(q_2^2)\sum_{\lam_{1,2}}
  \M_X(p_1,\sig_1;p_2,\sig_2;q_1,\lam_1;q_2,\lam_2) \nn\\
 &\quad\times\M_1(q_1,\lam_1;k_1;k_3)\,
             \M_2(q_2,\lam_2;k_2;k_4)
\label{amp}
\end{align}
with the $\tau$ propagator factor $D(q^2)=(q^2-m^2+im\Gamma)^{-1}$.

It is straightforward to obtain the squared matrix elements of the full
production plus decay amplitudes,
\begin{align}
 \sum|\M|^2 &\equiv
 \sum_{\sig_{1,2}}
 \big|\M(p_1,\sig_1;p_2,\sig_2;k_i)|^2 \nn\\
 &=\big|D(q_1^2)D(q_2^2)\big|^2
  \sum_{\lam_1,\lam_2}\sum_{\bar\lam_1,\bar\lam_2}
  {\p}^{\lam_1\lam_2}_{\bar\lam_1\bar\lam_2}\,
  {\D_1}^{\lam_1}_{\bar\lam_1}{\D_2}^{\lam_2}_{\bar\lam_2}
\label{amp2}
\end{align}
in terms of the production density matrix
${\p}^{\lam_1\lam_2}_{\bar\lam_1\bar\lam_2}$ and the decay density
matrices ${\D_{1,2}}^{\lam_{1,2}}_{\bar\lam_{1,2}}$; 
\begin{align}
 {\p}^{\lam_1\lam_2}_{\bar\lam_1\bar\lam_2}&=\sum_{\sigma}
  \M^{\lam_1\lam_2}_{\sig}
 \big(
  \M^{\bar\lam_1\bar\lam_2}_{\sig}
 \big)^*, \label{psdm}\\
 {\D_1}^{\lam_1}_{\bar\lam_1}&=
  \M_{\lam_1}
 \big(
  \M_{\bar\lam_1}
 \big)^*, \label{dsdm1}\\
 {\D_2}^{\lam_2}_{\bar\lam_2}&=
  \M_{\lam_2}
 \big(
  \M_{\bar\lam_2}
 \big)^*.\label{dsdm2}
\end{align}
In the narrow width limit the propagator factor becomes
\begin{align}
 |D(q^2)|^2\to \frac{\pi}{m\Gamma}\,\delta(q^2-m^2).
\end{align}
Although we only consider parton-level subprocesses, one can
generalize~\eqref{amp2} to mixed case and apply it for any processes,
including summation over different subprocesses and a product of the
relevant parton densities. We can also easily replace the
$\tau\to\nu_{\tau}\pi$ decay amplitude by one for the other hadronic
decay mode as well as the leptonic mode.

\subsection{Kinematics}

\begin{figure}
\center
 \includegraphics[width=.5\textwidth,clip]{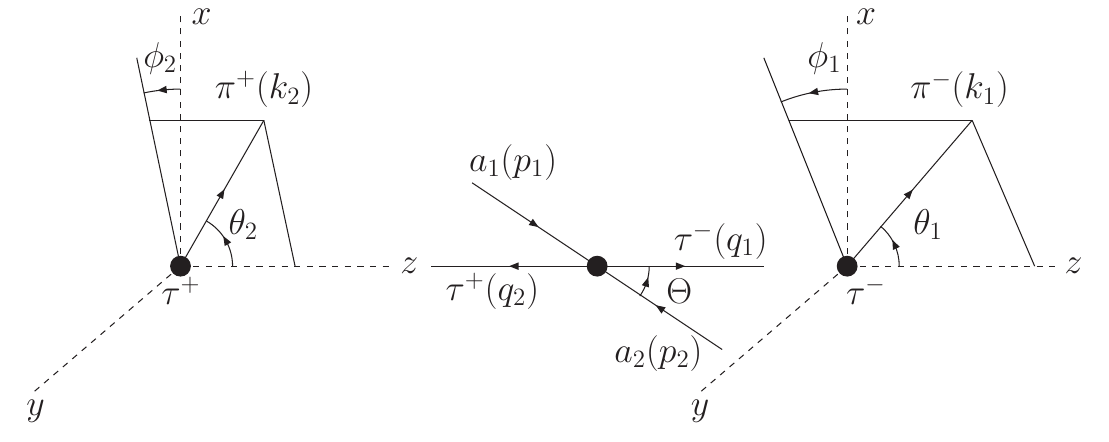}
 \caption{Schematic view of the coordinate system.}
\label{fig:frame}
\end{figure}

Let us define the kinematical variables. In the collision center-of-mass 
(CM) frame, i.e. in the $X$ rest frame, we choose the $\tau$ momentum
direction as the $z$-axis, 
\begin{align}
 p_1&=\tfrac{\sqrt{\hat s}}{2}(1,-\sin\Theta,0,\cos\Theta), \nn\\
 p_2&=\tfrac{\sqrt{\hat s}}{2}(1,\sin\Theta,0,-\cos\Theta), \nn\\
 q_1&=\tfrac{\sqrt{\hat s}}{2}\big(1+\tfrac{q_1^2-q_2^2}{\hat s},
   0,0,\beta\big), \nn\\
 q_2&=\tfrac{\sqrt{\hat s}}{2}\big(1+\tfrac{q_2^2-q_1^2}{\hat s},
   0,0,-\beta\big),
\end{align}
where
$\beta=\bar\beta\big(\frac{q_1^2}{\hat s},\frac{q_2^2}{\hat s}\big)$
with $\bar\beta(a,b)$ defined in~\eqref{beta}, $\Theta$ is the
scattering angle, and we choose $\vec p_1\times\vec q_1$ direction as
the $y$-axis. The momenta of the $\tm$ decay products are parametrized
in the $\tm$ rest frame,
\begin{align}
 k_1&=\tfrac{\sqrt{q_1^2}}{2}
  (1+\tfrac{m_{\pi}^2}{q_1^2},\beta_1\sin\theta_1\cos\phi_1,
       \beta_1\sin\theta_1\sin\phi_1,\beta_1\cos\theta_1), \nn\\
 k_3&=\tfrac{\sqrt{q_1^2}}{2}\beta_1
  (1,-\sin\theta_1\cos\phi_1,
   -\sin\theta_1\sin\phi_1,
   -\cos\theta_1),
\label{decay1}
\end{align}
with $\beta_1=1-m_{\pi}^2/q_1^2$. Similarly, those of the $\tp$ decay
products are 
\begin{align}
 k_2&=\tfrac{\sqrt{q_2^2}}{2}
  (1+\tfrac{m_{\pi}^2}{q_2^2},\beta_2\sin\theta_2\cos\phi_2,
       \beta_2\sin\theta_2\sin\phi_2,\beta_2\cos\theta_2), \nn\\
 k_4&=\tfrac{\sqrt{q_2^2}}{2}\beta_2
  (1,-\sin\theta_2\cos\phi_2,
   -\sin\theta_2\sin\phi_2,
   -\cos\theta_2),
\label{decay2}
\end{align}
with $\beta_2=1-m_{\pi}^2/q_2^2$. We note that the polar ($z$-)axis and
the $y$-axis normal to the scattering plane are chosen common to all the
three frames, and the two decay frames differ only by the boost along
the $\tau$ production axis (see Fig.~\ref{fig:frame}). The $\tau$ width
is very narrow, $\Gamma\sim{\cal O}(10^{-12}\,{\rm GeV})$, and hence we
take the narrow width limit, $q_1^2=q_2^2=m^2$, in the following
analytic amplitudes. 

\subsection{Tau-pair production and decay amplitudes}

The helicity amplitude for the tau-pair production via
$Z$-boson~\eqref{processZ} is given by 
\begin{align}
 \M_Z=\frac{e^2\hat s}{\hat s-m_Z^2+im_Z\Gamma_Z}
  \hat\M_{\sig}^{\lam_1\lam_2},
\end{align}
where the reduced amplitude is
\begin{align}
 &\hat\M_{\sig}^{\lam_1\lam_2} \nn\\
 &=\begin{cases}
   -\frac{1}{2}
    g_{\sig}^q\{g_+^{\tau}(1+\lam_1\beta)+g_-^{\tau}(1-\lam_1\beta)\}
    (\sig\lam_1+\cos\Theta) \\
    \hspace*{5.5cm}\text{for }\lam_1\ne\lam_2, \\
    \frac{m}{\sqrt{\hat s}}
     \lam_1g_{\sig}^q(g_+^{\tau}+g_-^{\tau})\sin\Theta
    \hspace*{2cm}\text{for }\lam_1=\lam_2.
  \end{cases}
\end{align}
Here, we neglect the initial fermion mass and take
$\sig=\sig_1=-\sig_2$. $g_{\pm}^q$ and $g_{\pm}^{\tau}$ are the
$Z$-boson couplings to right- and left-handed quarks and tau leptons,
respectively. 

The amplitude for the $S(=H,A)$-boson case~\eqref{processH} and
\eqref{processA} is 
\begin{align}
 \M_S=\frac{-y_{\tau}g_{S}\hat s}{\hat s-m_{S}^2+im_{S}\Gamma_{S}}
 \frac{\sqrt{\hat s}}{2}\, \hat\M_{\sig}^{\lam_1\lam_2},
\end{align}
where
\begin{align}
 \hat\M_{\sig}^{\lam\lam} &= \beta\lam & \text{for }S=H, \\
 \hat\M_{\sig}^{\lam\lam} &= i & \text{for }S=A.
\end{align}
The amplitudes are non-zero only when $\sig=\sig_1=\sig_2$ and
$\lam=\lam_1=\lam_2$. Here, $y_{\tau}=\sqrt{2}m_{\tau}/v$ is the $\tau$
Yukawa coupling, and $g_S$ is the $ggS$ coupling, which is given by
$g_H=\alpha_s g_{Htt}/3\pi v$ and $g_A=\alpha_s g_{Att}/2\pi v$ in the
heavy-top limit. 

The $\tau\to\pi\nu$ decay amplitudes in the $\tau$ rest frame are
\begin{align}
 \M_{1,2}=-G_Ff_{\pi}m^2\sqrt{\beta_{1,2}}\,\hat\M_{\lam_{1,2}},
\end{align}
where
\begin{subequations}
\label{decayamp}
\begin{align}
 \hat\M_{\lam_1}&=\sqrt{1+\lam_1\cos\theta_1}\,e^{i\lam_1\phi_1/2}, \\
 \hat\M_{\lam_2}&=\sqrt{1+\lam_2\cos\theta_2}\,e^{-i\lam_2\phi_2/2}.
\end{align}
\end{subequations}
Those amplitudes are invariant with the boost to the collision CM
frame.

\subsection{Helicity correlations}

The trivial helicity correlations are given by the diagonal parts of the 
density matrices, $\lam_{1}=\bar\lam_{1}$ and $\lam_{2}=\bar\lam_{2}$.
In the following analytic expressions we take 
$m_{\tau}/\sqrt{\hat s}=0$, i.e. $\beta=1$.

\begin{figure}
\center
 \includegraphics[width=.24\textwidth,clip]{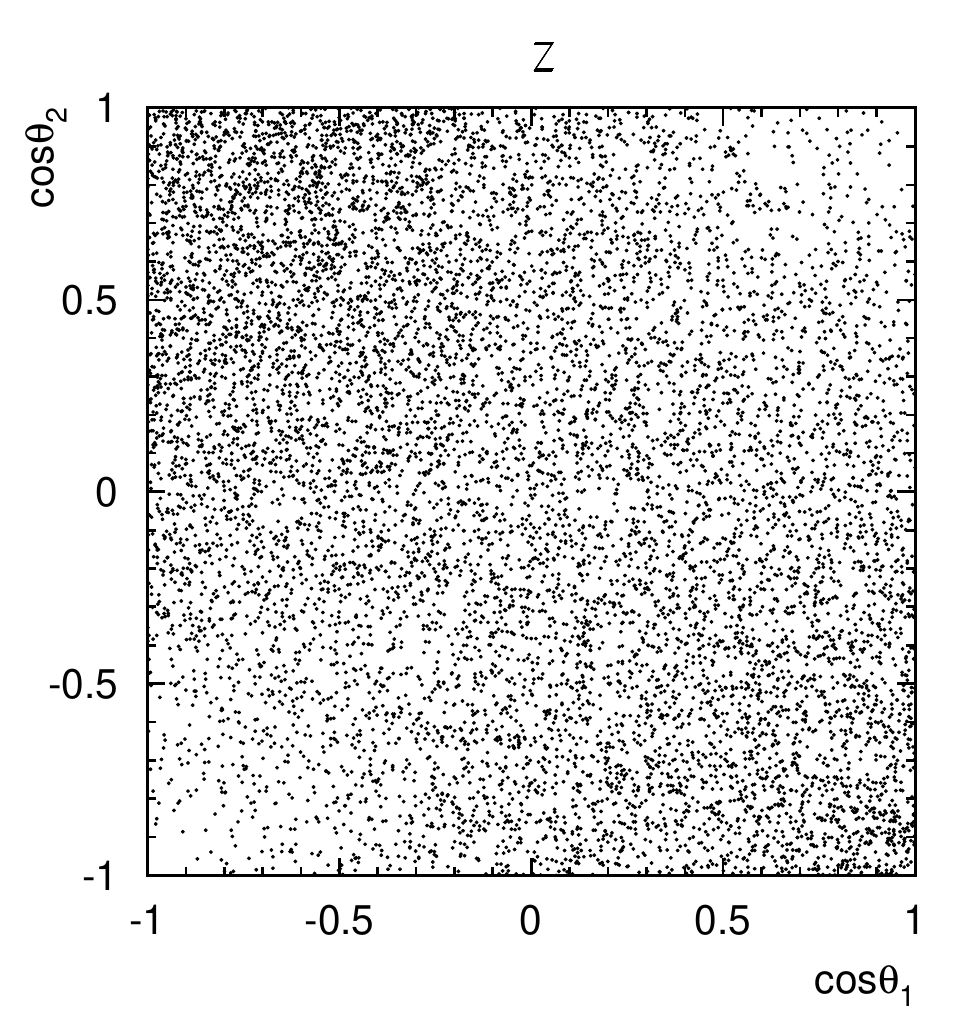}
 \includegraphics[width=.24\textwidth,clip]{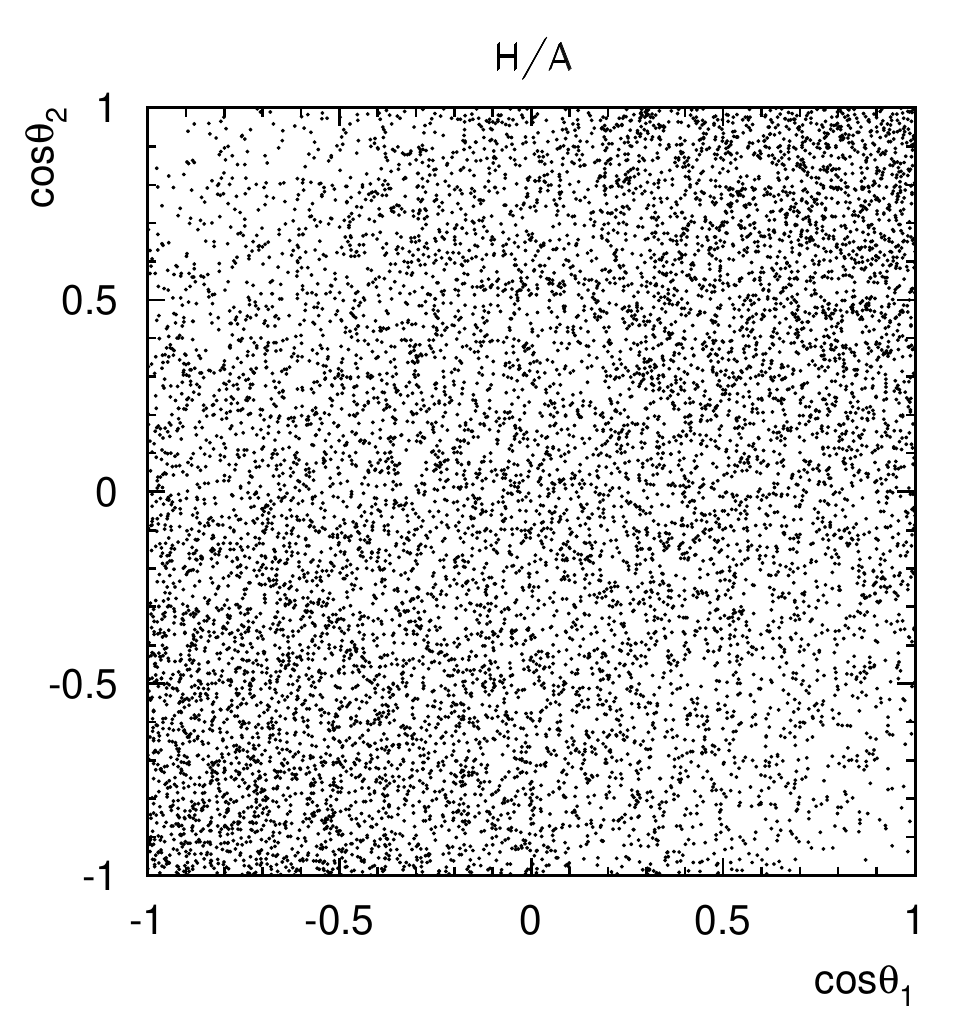}
 \caption{$\cos\theta_1$-$\cos\theta_2$ correlation in
 $pp\to X\to\tau^-(\to\pi^-\nu)\,\tau^+(\to\pi^+\bar\nu)$ for $X=Z$
 (left) and $H/A$ (right). The pion polar angles $\cos\theta_{1,2}$ are
 defined in the $\tau$ rest frame; see Fig.~\ref{fig:frame}.}
\label{fig:cos1cos2}
\end{figure}

For the $Z$ production, only for $\lam_1\ne\lam_2$ the production
amplitude is non-zero. After integrating out the scattering angle
$\Theta$ and the azimuthal angles $\phi_{1,2}$, the squared matrix
element~\eqref{amp2} is
\begin{align}
  {\p}^{+-}_{+-}&{\D_1}^+_+{\D_2}^-_-
 +{\p}^{-+}_{-+}{\D_1}^-_-{\D_2}^+_+ \nn\\
 &\propto (1-\cos\theta_1\cos\theta_2)
 +\kappa\,(\cos\theta_1-\cos\theta_2).
\end{align}
with
$\kappa=({g_+^{\tau}}^2-{g_-^{\tau}}^2)/({g_+^{\tau}}^2+{g_-^{\tau}}^2)\sim
-0.15$.
As seen in \eqref{decayamp}, the polar angle $\cos\theta_{1,2}$
distribution of the pion arising from $\tau^{\pm}_L$ is
$(1-\cos\theta_{1,2})$, while from $\tau^{\pm}_R$ it is
$(1+\cos\theta_{1,2})$. Hence in $Z\to\tp_L\tm_R$ or $\tp_R\tm_L$ decays
the two pions tend to be emitted to the opposite direction along the
$z$-axis. The difference between the left and right coupling,
i.e. parity violation, gives a small linear dependence of $\cos\theta_1$
and $\cos\theta_2$ in the single differential cross section, and one can
see a slightly higher density in the $\cos\theta_1<0$ and
$\cos\theta_2>0$ region in the left panel of Fig.~\ref{fig:cos1cos2}. 

For the Higgs production, on the other hand, only for $\lam_1=\lam_2$
the production amplitude is non-zero, and therefore
\begin{align}
  {\p}^{++}_{++}{\D_1}^+_+{\D_2}^+_+
 +{\p}^{--}_{--}{\D_1}^-_-{\D_2}^-_-
 \propto (1+\cos\theta_1\cos\theta_2).
\end{align}
Apart from the linear dependence of $\cos\theta_{1,2}$ for the $Z$ case, 
completely the opposite is favoured for $H/A\to\tp_L\tm_L$ or
$\tp_R\tm_R$ decays as shown in the right panel of
Fig.~\ref{fig:cos1cos2}. Note that the helicity correlations for $H$ and
$A$ are identical. For comparison the masses of $H$ and $A$ are assumed
to be the $Z$-boson mass, $m_{H/A}=m_Z$.

\subsection{Polarization correlations}

The non-trivial polarization correlations are given by the off-diagonal 
parts of the density matrices, i.e. $\lam_{1}=-\bar\lam_{1}$ and
$\lam_{2}=-\bar\lam_{2}$, which produce the azimuthal angle
dependence. When we isolate the azimuthal angle dependence in
(\ref{amp2}), there are nine distributions (including one constant
piece) as 
\begin{align}
 &{\p}^{\lam_1\lam_2}_{\bar\lam_1\bar\lam_2}\,
  {\D_1}^{\lam_1}_{\bar\lam_1}{\D_2}^{\lam_2}_{\bar\lam_2}  \nn\\
 &= F_1^{}+\big\{2\,{\Re}e\big[F_2^{}\cos\phi_1+F_3^{}\cos\phi_2 \nn\\
  &\hspace*{2cm}+F_4^{+}\cos(\phi_1+\phi_2)
                +F_4^{-}\cos(\phi_1-\phi_2)\big] \nn\\
  &\hspace*{1.5cm}+({\Re}e\to{\Im}m,\,\cos\to\sin)\big\}.
\label{general_azdis}
\end{align}
Here, and in the following, summation over repeated indices
$(\lam_1,\lam_2,\bar\lam_1,\bar\lam_2)=\pm$ is implied. The coefficients
$F_i^{(\pm)}$ are the functions of the kinematical variables except the
azimuthal angles $\phi_{1,2}$. For the production of a $Z$ boson, they
also depend on the production angle $\Theta$. For the spin-0 particle
case, only the $F_4^-$ term in (\ref{general_azdis}) survives due to the
helicity selection $\lam_1=\lam_2$. All the {\it sine} terms vanish when
$CP$ is conserved and when the absorptive part of the amplitudes are
neglected, {\it e.g.}, in the tree-level approximation.

Because the phase of the product of the two decay density matrices is 
\begin{align}
 {\D_1}^{\lam_1}_{\bar\lam_1}{\D_2}^{\lam_2}_{\bar\lam_2}\propto
 \exp[i\{(\lam_1-\bar\lam_1)\phi_1-(\lam_2-\bar\lam_2)\phi_2\}/2],
\label{J1J2phase}
\end{align}
the coefficients $F_{1-4}^{(\pm)}$ are expressed in terms of the
production density matrix and two decay density matrices as
\begin{align}
 F_1^{}&=\p^{\lam_1\lam_2}_{\lam_1\lam_2}
  {\D_1}^{\lam_1}_{\lam_1}{\D_2}^{\lam_2}_{\lam_2}, \nn \\
 F_2^{}&=\p^{+\lam_2}_{-\lam_2}
  {\D_1}^{+}_{-}{\D_2}^{\lam_2}_{\lam_2}, \nn \\
 F_3^{}&=\p^{\lam_1+}_{\lam_1-}
  {\D_1}^{\lam_1}_{\lam_1}{\D_2}^{+}_{-}, \nn \\
 F_4^{\pm}&=\p^{+\mp}_{-\pm}
  {\D_1}^{+}_{-}{\D_2}^{\mp}_{\pm}.
\label{F1to9}
\end{align}
The azimuthal angle correlations are manifestly expressed by quantum
interference among different helicity states of the intermediate tau
leptons. 

\begin{figure}
\center
 \includegraphics[width=0.24\textwidth,clip]{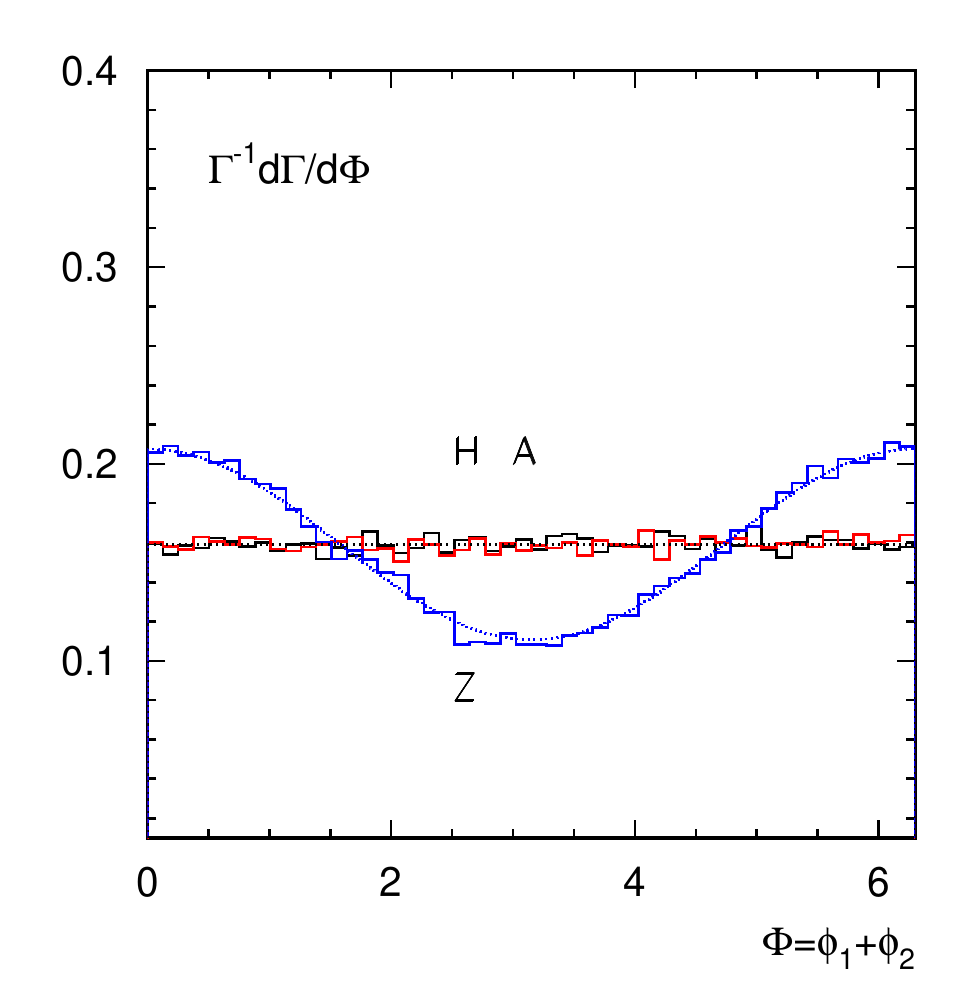}
 \includegraphics[width=0.24\textwidth,clip]{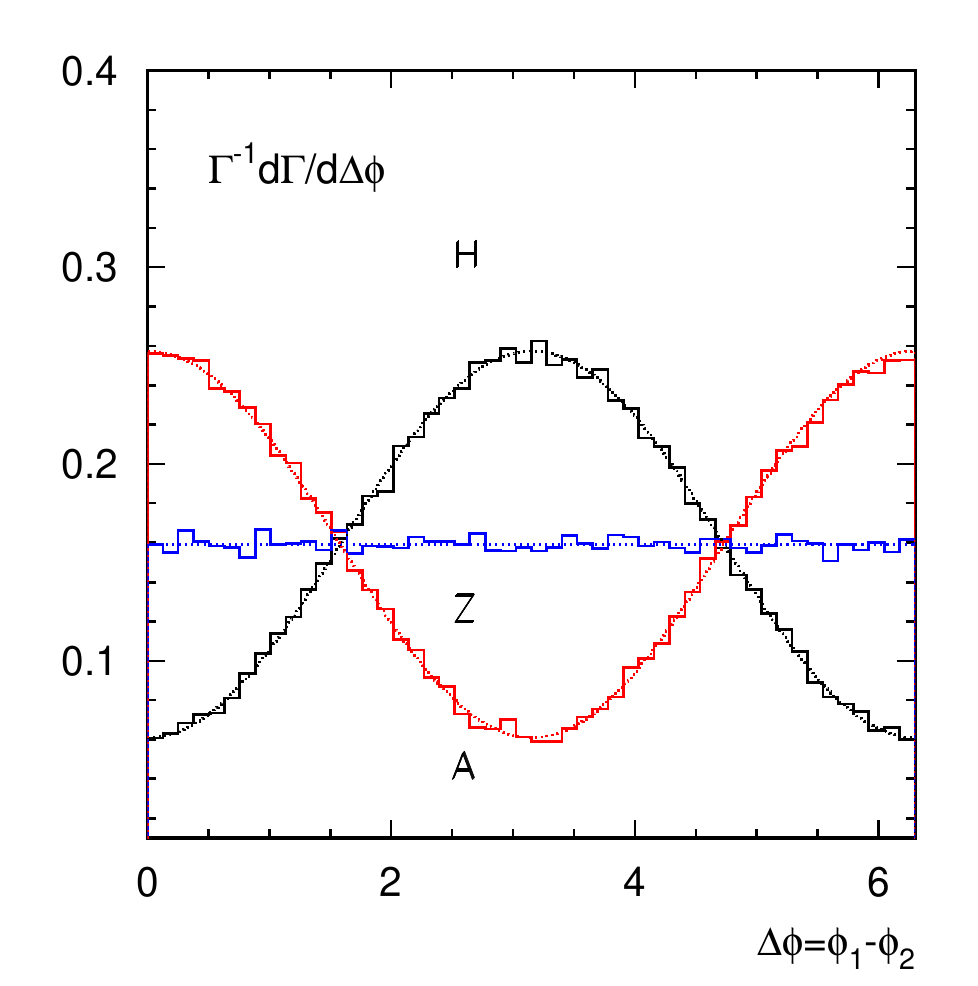}
 \caption{Azimuthal angle correlations, $\phi_1+\phi_2$ (left) and
 $\phi_1-\phi_2$ (right), in 
 $pp\to X\to\tau^-(\to\pi^-\nu)\,\tau^+(\to\pi^+\bar\nu)$ for $X=Z$
 (blue), $H$ (black), and $A$ (red). The pion
 azimuthal angles $\phi_{1,2}$ are defined in the $\tau$ rest frame by
 the $pp\to\tau^+\tau^-$ scattering plane; see Fig.~\ref{fig:frame}.}
\label{fig:dphipphi}
\end{figure}

The $Z$ production could produce the $\phi_1$ (or $\phi_2$) dependence, 
however this is very small because $F_2$ ($F_3$) is proportional to
$m/\sqrt{s}$. Similarly, $F_4^-\propto {\cal O}(m^2/\hat s)$, and hence
the distribution of the azimuthal difference between the two
$\tau$-decay planes, $\phi_1-\phi_2\,(\equiv\Delta\phi)$, is flat. An
interesting correlation between the two decay planes for the $Z$ case is
the $\phi_1+\phi_2\,(\equiv\Phi)$ correlation, whose coefficient $F_4^+$
is 
\begin{align}
 F_4^+ \propto -g_+^{\tau}g_-^{\tau}
   \sin^2\Theta\sin\theta_1\sin\theta_2.
\end{align}
After integrating out $\Theta$, $\theta_1$ and $\theta_2$, the azimuthal
asymmetry is given by 
\begin{align}
  \frac{1}{\Gamma}\frac{d\Gamma}{d\Phi}
 =\frac{1}{2\pi}\big[1+A_4^+\cos\Phi\big]
\end{align}
with
\begin{align}
 A_4^+ \equiv \frac{2F_4^+}{F_1}
 =\frac{\pi^2}{16}
   \frac{-g_+^{\tau}g_-^{\tau}}{{g_+^{\tau}}^2+{g_-^{\tau}}^2}
 \sim 0.30.
\end{align}
The distribution is enhanced around $\Phi=0$ and $2\pi$, while it is
suppressed around $\Phi=\pi$.

For the $H/A$ production, as mentioned above, only the $\phi_1-\phi_2$
term is non-zero and the coefficient is
\begin{align}
 F_4^- \propto \mp\sin\theta_1\sin\theta_2,
\end{align}
where the $-/+$ sign is for $CP$-even/odd scalar. The azimuthal
asymmetry is given by~\cite{Kramer:1993jn} 
\begin{align}
  \frac{1}{\Gamma}\frac{d\Gamma}{d\Delta\phi}
 =\frac{1}{2\pi}\big[1+A_4^-\cos\Delta\phi\big]
\end{align}
with
\begin{align}
 A_4^- \equiv \frac{2F_4^-}{F_1}
 =\mp\frac{\pi^2}{16}
 \sim \mp 0.62.
\end{align}
The $CP$-even and -odd scalars have opposite modulation, and the
distribution is strongly enhanced (suppressed) around $\Delta\phi=\pi$
for the $CP$-even (-odd) scalars. 

To examine the validity of the model file, Fig.~\ref{fig:dphipphi} shows
the normalized azimuthal correlations between $\pi^-$ and $\pi^+$, and
our numerical results (solid histograms) agree well with the above
analytic formula (dotted lines).

\subsection{Spin correlations at the LHC}

\begin{figure*}
\center
 \includegraphics[width=.245\textwidth,clip]{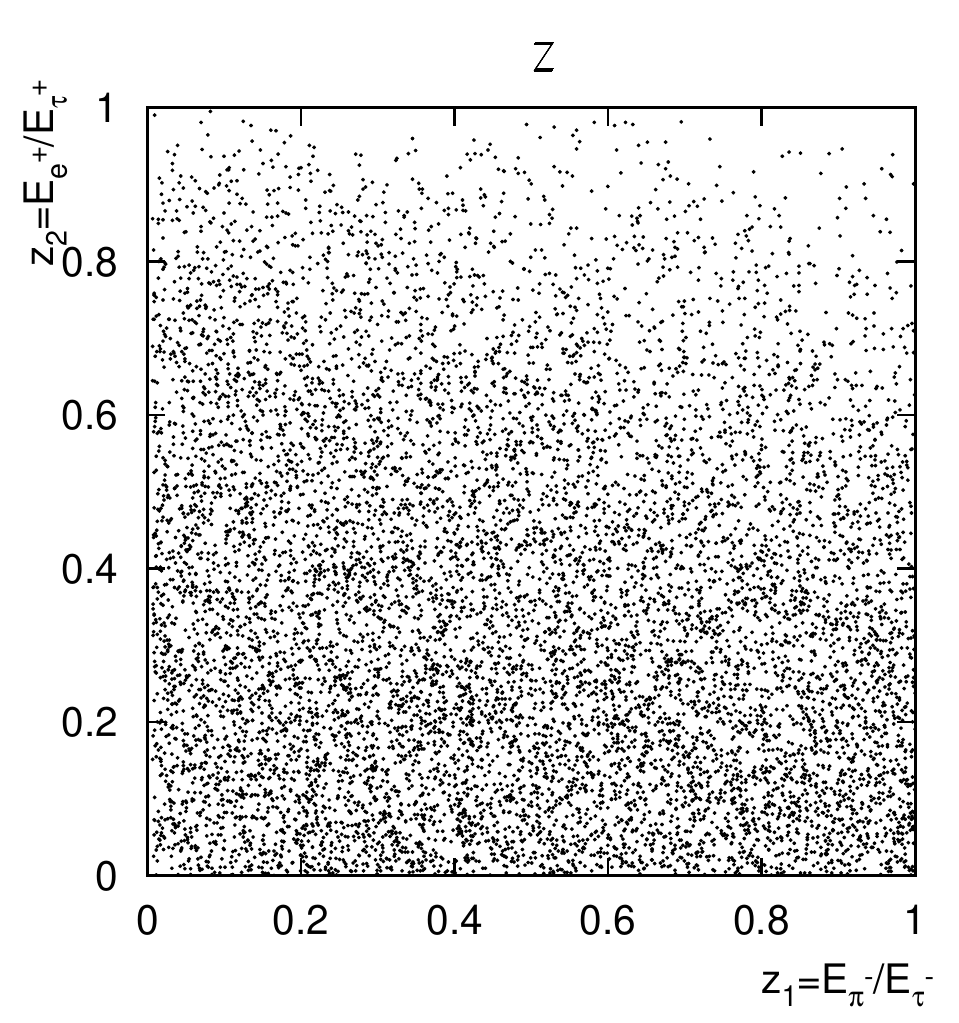}
 \includegraphics[width=.245\textwidth,clip]{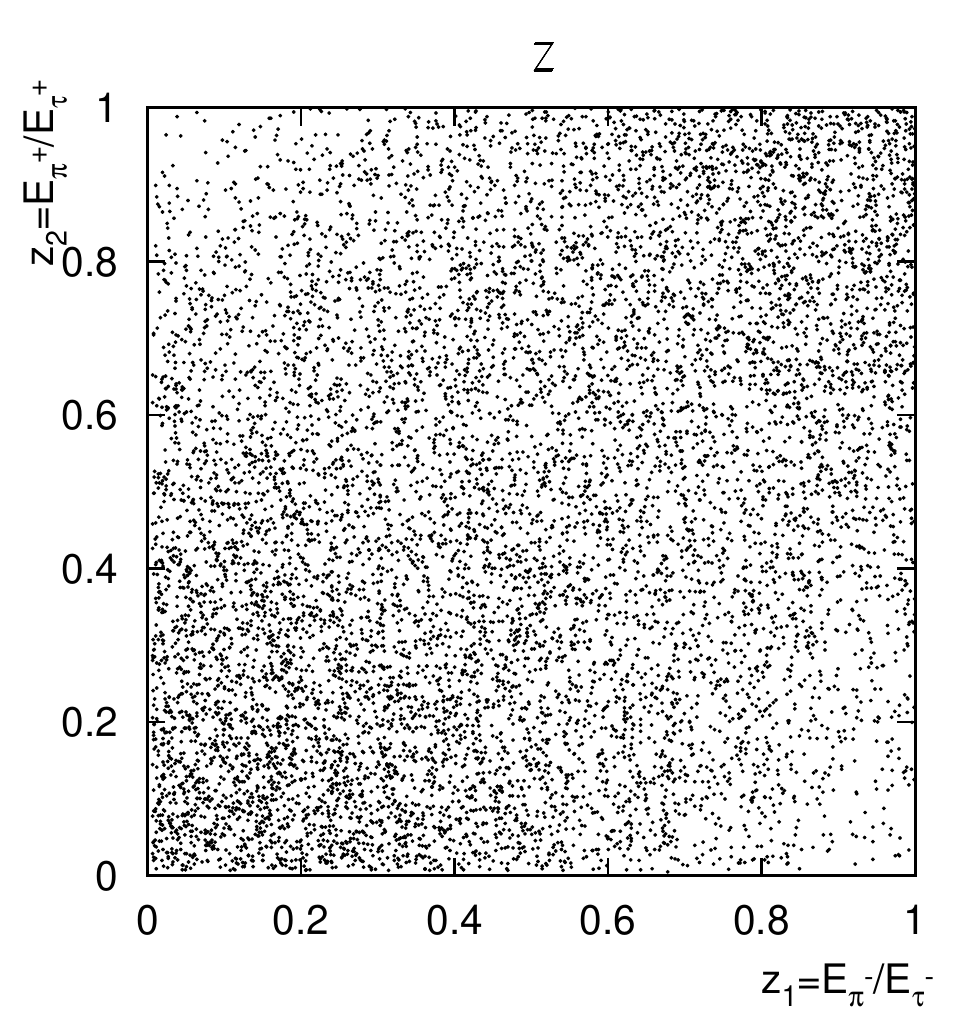}
 \includegraphics[width=.245\textwidth,clip]{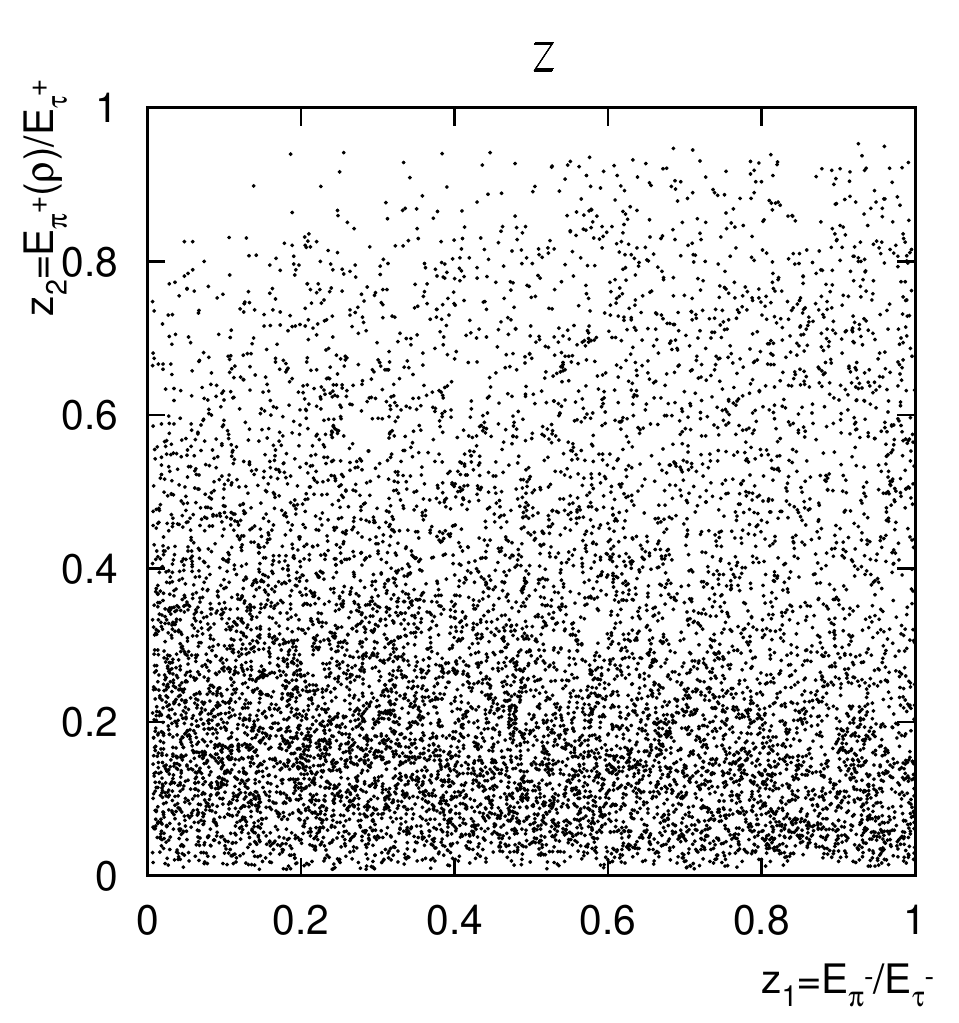}
 \includegraphics[width=.245\textwidth,clip]{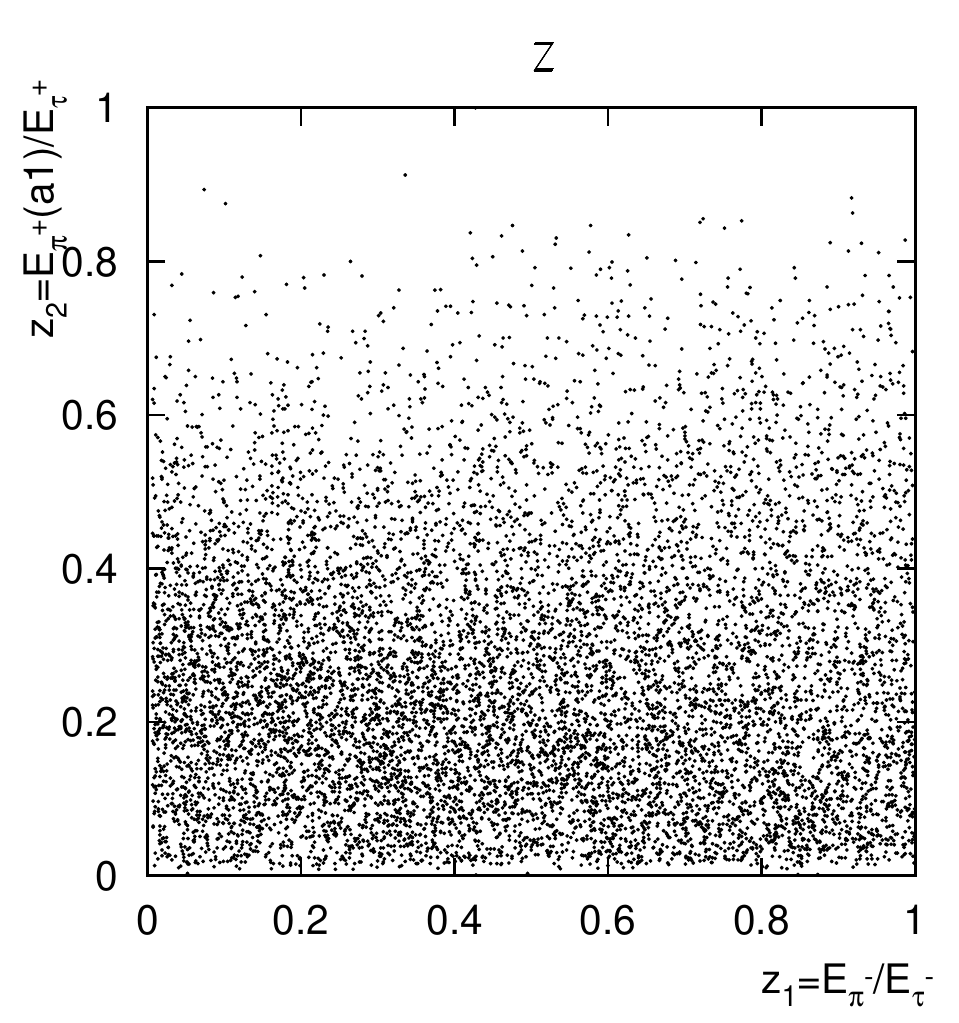}
 \includegraphics[width=.245\textwidth,clip]{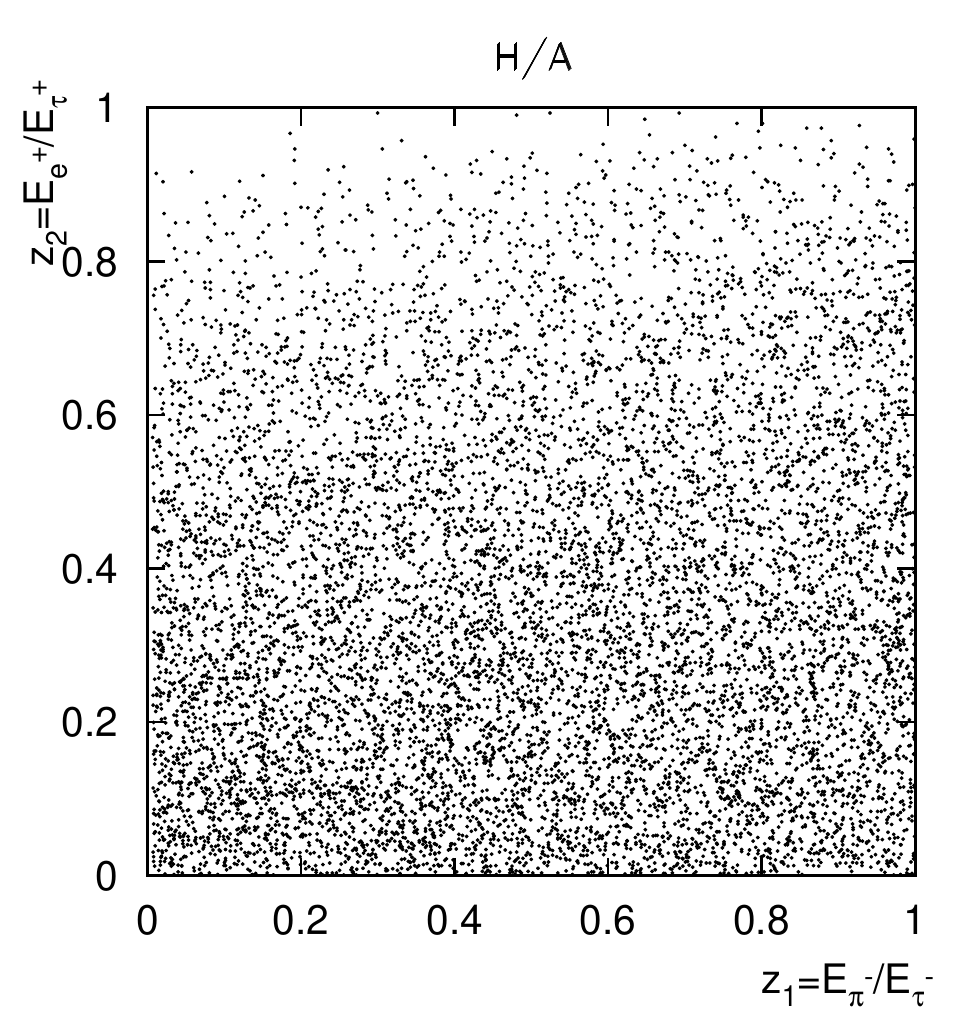}
 \includegraphics[width=.245\textwidth,clip]{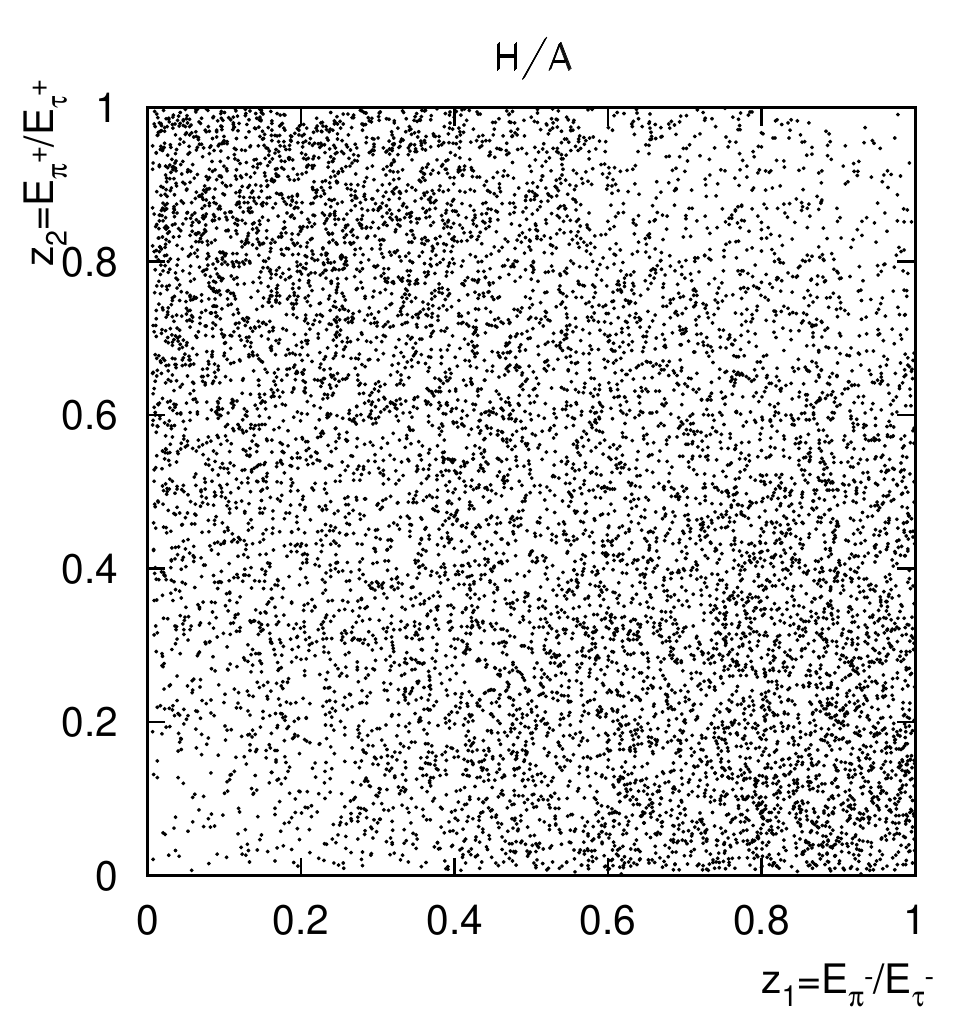}
 \includegraphics[width=.245\textwidth,clip]{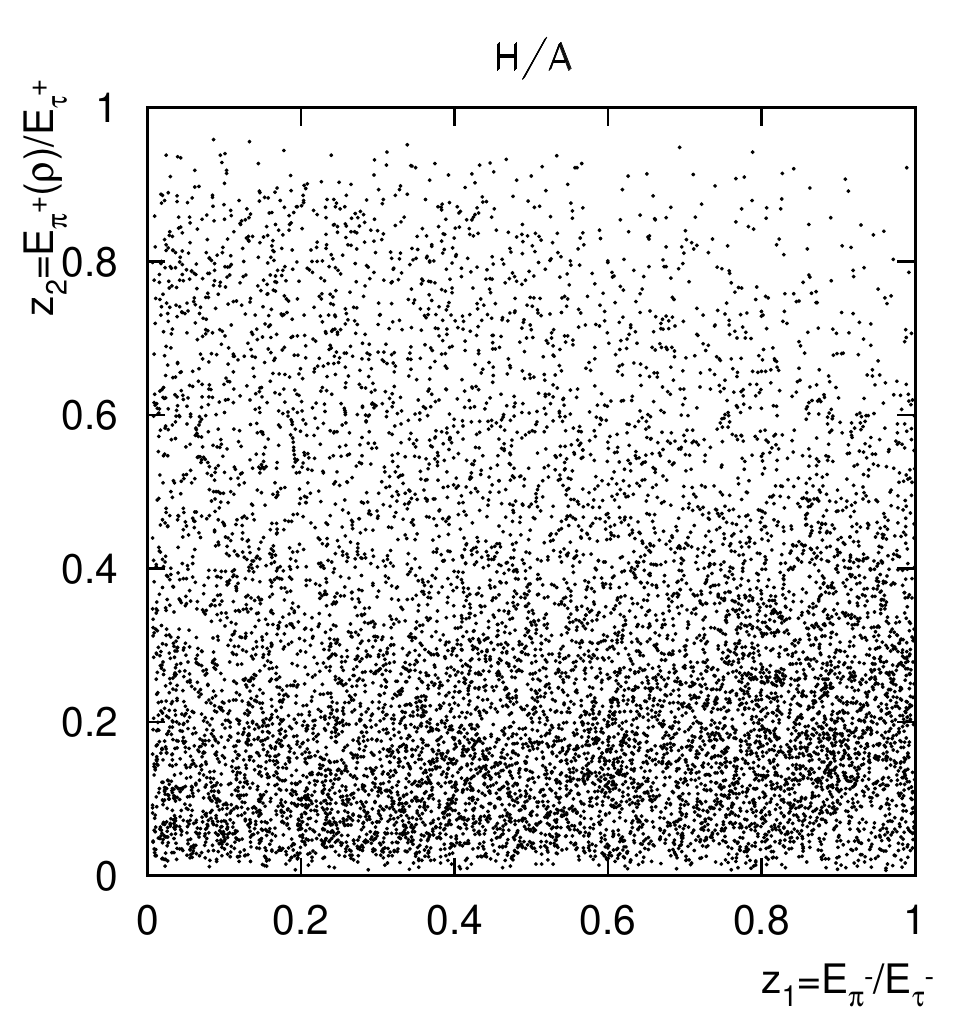}
 \includegraphics[width=.245\textwidth,clip]{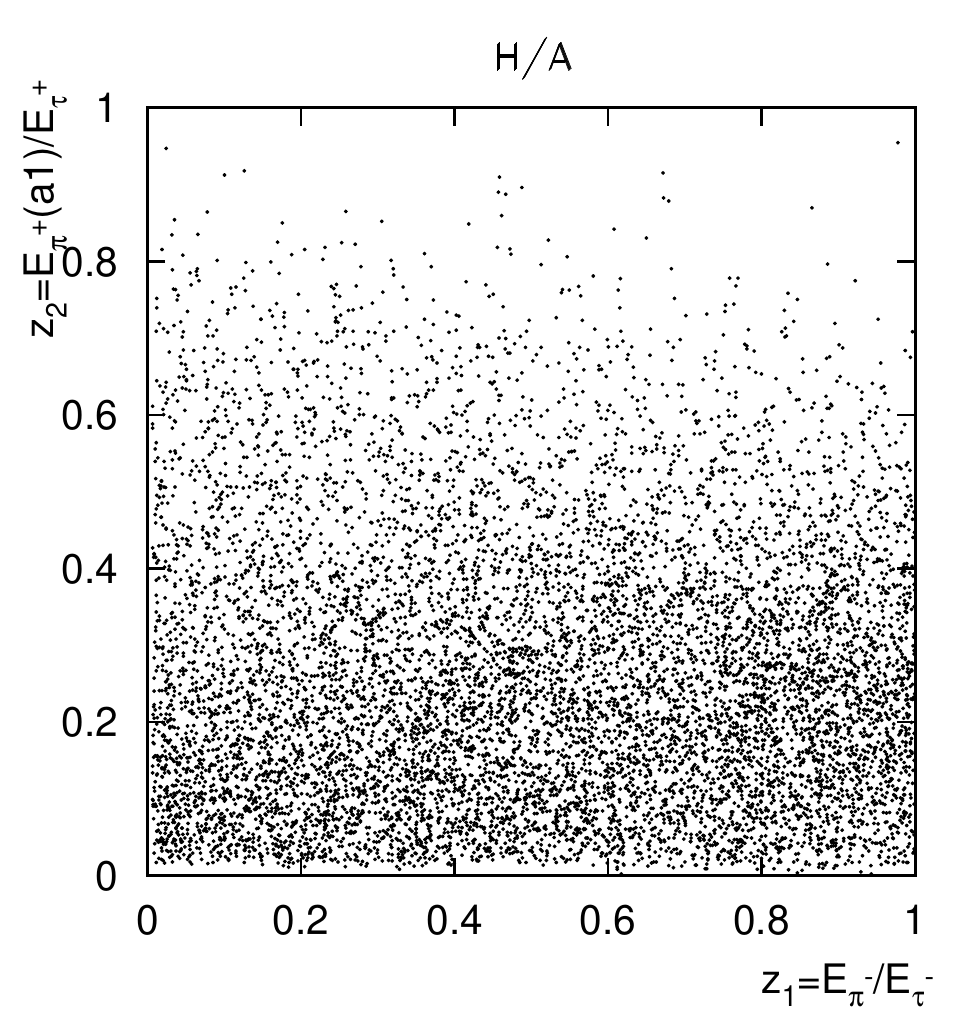}
 \caption{The fractional energy correlations in the laboratory frame in 
 $pp\to X\to\tau^-(\to\pi^-\nu)\,\tau^+(\to A^+\bar\nu)$ at
 $\sqrt{s}=8$~TeV for $X=Z$ (top) and  $H/A$ (bottom), where $A^+$ is
 the leptonic, $\pi$, $\rho$, and $a_1$ decay modes from the left to
 right. The resonant mass is taken to be the $Z$-boson mass.}
\label{fig:z1z2}
\end{figure*}

To present the validation of our program, we have so far discussed the
angular distributions, assuming that all the $\tau$-decay products can
be reconstructed. Here, we show some distributions of the realistic
experimental observables at the LHC.

First, we extend the helicity correlation in $\tau\to\nu_\tau\pi$
discussed above to other tau decay modes. In Fig.~\ref{fig:z1z2} we
present the helicity correlations in terms of energy fractions in the
laboratory frame for 
$pp\to Z/H/A\to\tau^-(\to \pi^-\nu)\,\tau^+(\to A^+\bar\nu)$, where
$A^+$ is the leptonic, $\pi$, $\rho$, and $a_1$ decay modes,
respectively, from left to right. The one-prong $\tau$ decays are
considered and the energy fractions are $z_1=E_{\pi^-}/E_{\tau^-}$ and
$z_2=E_{\hat A^+}/E_{\tau^+}$ with
$\hat A=e,\pi,\pi(\rho),\pi(a_1)$. For comparison, the resonant mass is
taken to the $Z$-boson mass, and hence the both taus are highly boosted,
where the collinear approximation can be safely applied. For the $A=\pi$
mode, in the collinear limit ($\beta=1$) we have
\begin{align}
 z_1=(1+\cos\theta_1)/2\quad\text{and}\quad
 z_2=(1-\cos\theta_2)/2.
\end{align}
As a result, the $z_1$-$z_2$ correlation behaves opposite to
$\cos\theta_1$-$\cos\theta_2$ correlation as shown in
Fig.~\ref{fig:cos1cos2}. 
The helicity correlations are different between the spin-1 and spin-0
bosons, while those are identical between the $CP$-even and -odd
scalars. 

\begin{figure}
\center
 \includegraphics[width=0.24\textwidth,clip]{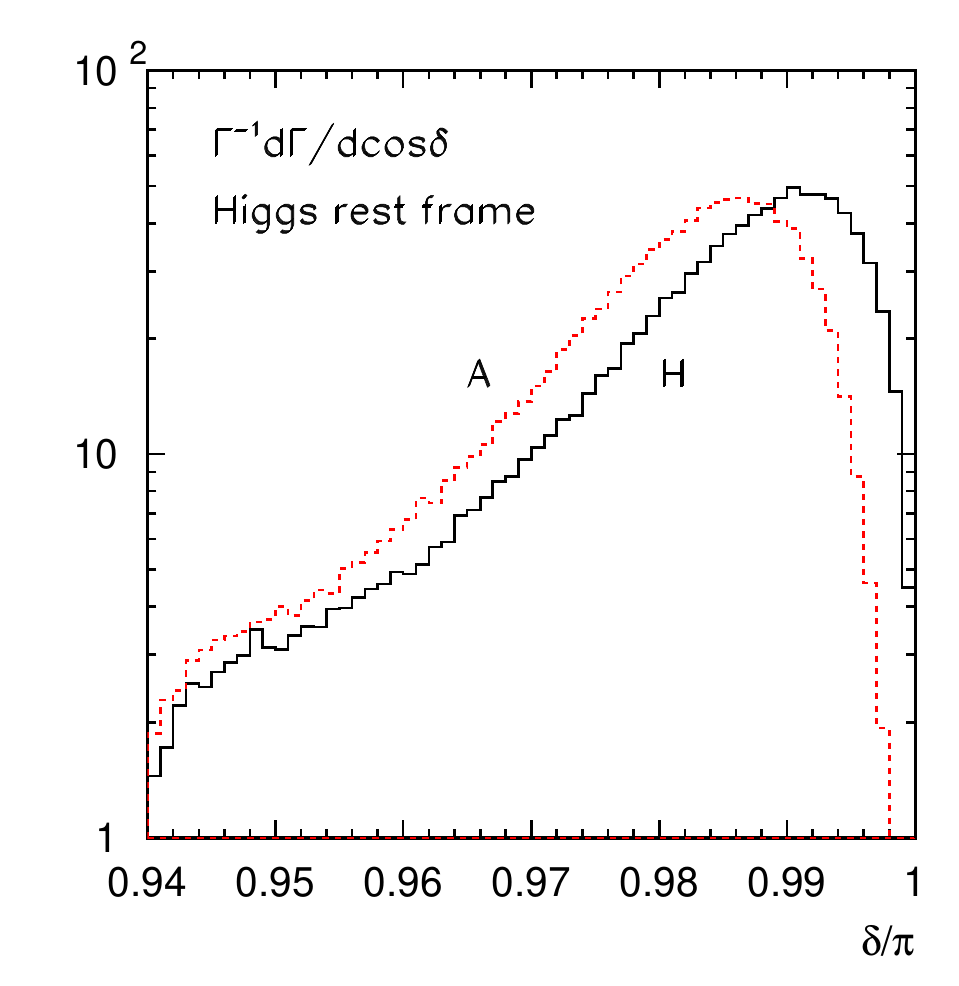}
 \includegraphics[width=0.24\textwidth,clip]{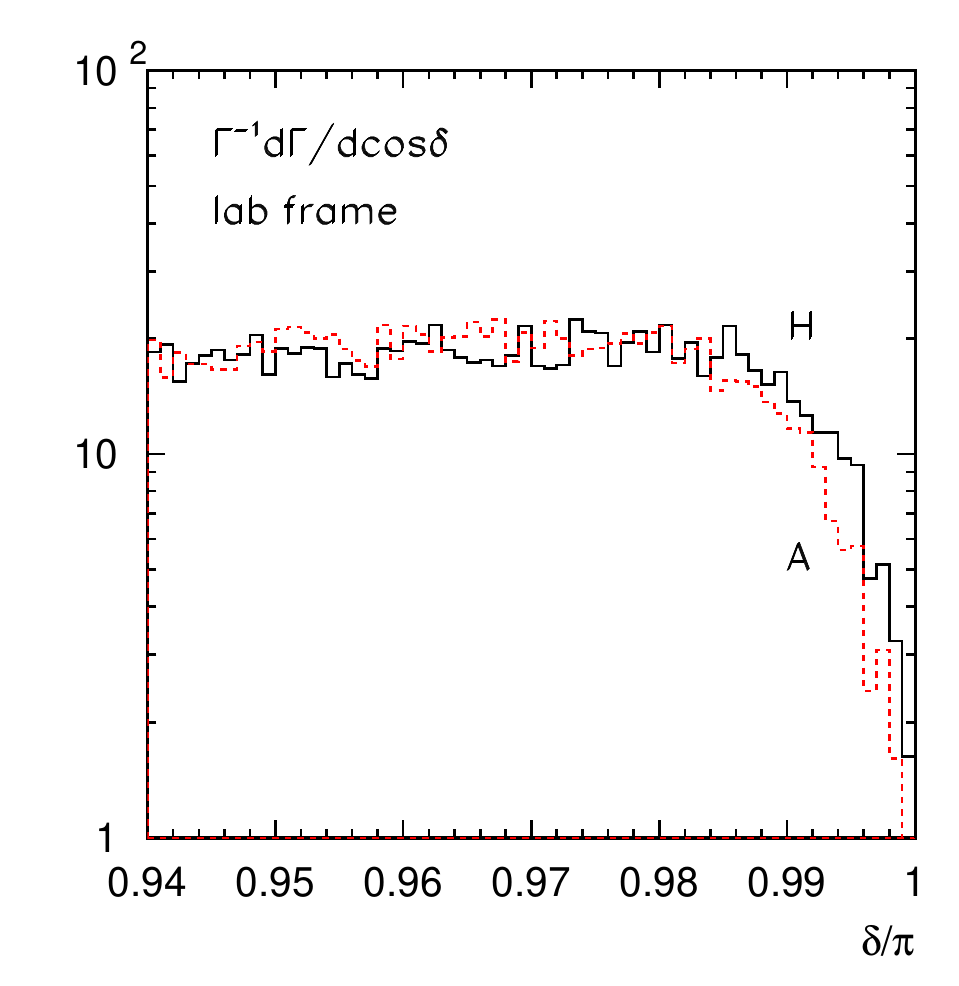}
 \caption{The $\pi^+\pi^-$ acollinearity distribution in the Higgs rest 
 frame (left) and in the laboratory frame (right) for
 $pp\to H/A\to\tau^-(\to \pi^-\nu)\,\tau^+(\to \pi^+\bar\nu)$ with
 $m_{H/A}=125$~GeV at $\sqrt{s}=8$~TeV.} 
\label{fig:delta}
\end{figure}

Second, an useful observable to measure the parity of Higgs boson is the
acollinearity angle, namely the angle $\delta$ between $\pi^+$ and
$\pi^-$~\cite{Kramer:1993jn}. The azimuthal asymmetry $\Delta\phi$ can
be given as a function of angle $\delta$. To display the measurable
difference of Higgs parities at the LHC, in Fig.~\ref{fig:delta}, we
show the $\pi^+\pi^-$ acollinearity distribution in the Higgs rest frame
(left) and in the lab frame (right) for 
$pp\to H/A\to\tau^-(\to\pi^-\nu)\,\tau^+(\to \pi^+\bar\nu)$ at
$\sqrt{s}=8$~TeV. Here we take the Higgs mass as 125~GeV. One can see
the distinguishable difference between $CP$-even and $CP$-odd scalars
around $\delta=\pi$, although in the lab frame the boost effect leads to
relatively unclear discrepancy. While the reconstruction of the Higgs
rest frame at hadron colliers is difficult, it is possible in the
$Z$-associated Higgs production at $e^+e^-$ colliders and the
acollinearity distribution in the reconstructed $H/A$ rest frame has
been discussed in~\cite{Was:2002gv}.

\section{Spin correlations with {\sc TauDecay}}\label{sec:taudecay}

To validate our {\sc FR} $\tau$-decay model file, all the results shown
in Sections~\ref{sec:stau} and \ref{sec:spin} are generated in {\sc MG5}
with intermediate taus being as propagators, denoted by {\it
method~A}. Unless intermediate taus are off-shell such as the case in
Sect.~\ref{sec:stau}, a library which collects all the possible $\tau$
decay channels would be much more efficient for practical event
generations since $\tau$ production and its decay can be simulated
independently as we use {\sc Tauola}. 

{\sc TauDecay} is a {\sc Fortran} library of $\tau$-decay helicity
amplitudes, constructed in the FR/MG5 framework as described in
Sect.~\ref{sec:fr}.  The input of the subroutines for each decay channel
is the four momenta of the decaying $\tau$ as well as its decay products
and their helicities. The output is the amplitude, i.e. a complex
number, at a given phase space point with a given helicity
configuration. In this section, we present how the {\sc TauDecay}
package can reproduce spin correlation discussed in
Sect.~\ref{sec:spin}. 

We show two methods by means of {\sc TauDecay}; without and with spin
density matrix, denoted by {\it method~B} and {\it C}, respectively. The
both methods are carefully checked by {\it method A}.
\begin{itemize}
\item[{\it B}:] ({\it without density matrix}) The $\tau$ production
	     amplitude as well as the decay amplitudes from {\sc
	     TauDecay} are evaluated for a given process for a given
	     phase space point. The $\tau$ helicities in the product of
	     the production and decay amplitudes should be summed over
	     as indicated in~\eqref{amp}, and then the $\tau$ helicity
	     summed amplitudes should be squared. This is theoretically
	     identical to the propagator method in the narrow width
	     limit, but it is useful for systematically studying
	     different tau decay modes for each production process.  
\item[{\it C}:] ({\it with density matrix}) First, the $\tau$ production
	     density matrix is computed for each phase space point. If
	     there is only one tau in the final state, this will be a
	     $2\times2$ matrix, while it will be a $4\times4$ double
	     density matrix for two taus as
	     ${\p}^{\lam_1\lam_2}_{\bar\lam_1\bar\lam_2}$
	     in~\eqref{psdm}. Then, the correlated tau decays are
	     simulated by multiplying the production density matrix with
	     the corresponding $\tau$ decay helicity amplitudes and
	     their complex conjugates provided by {\sc TauDecay}, namely
	     the decay density matrices
	     ${\D_{i}}^{\lam_{i}}_{\bar\lam_{i}}$, according
	     to~\eqref{amp2}. 
\end{itemize}

\begin{figure}
\center
 \includegraphics[width=.35\textwidth,clip]{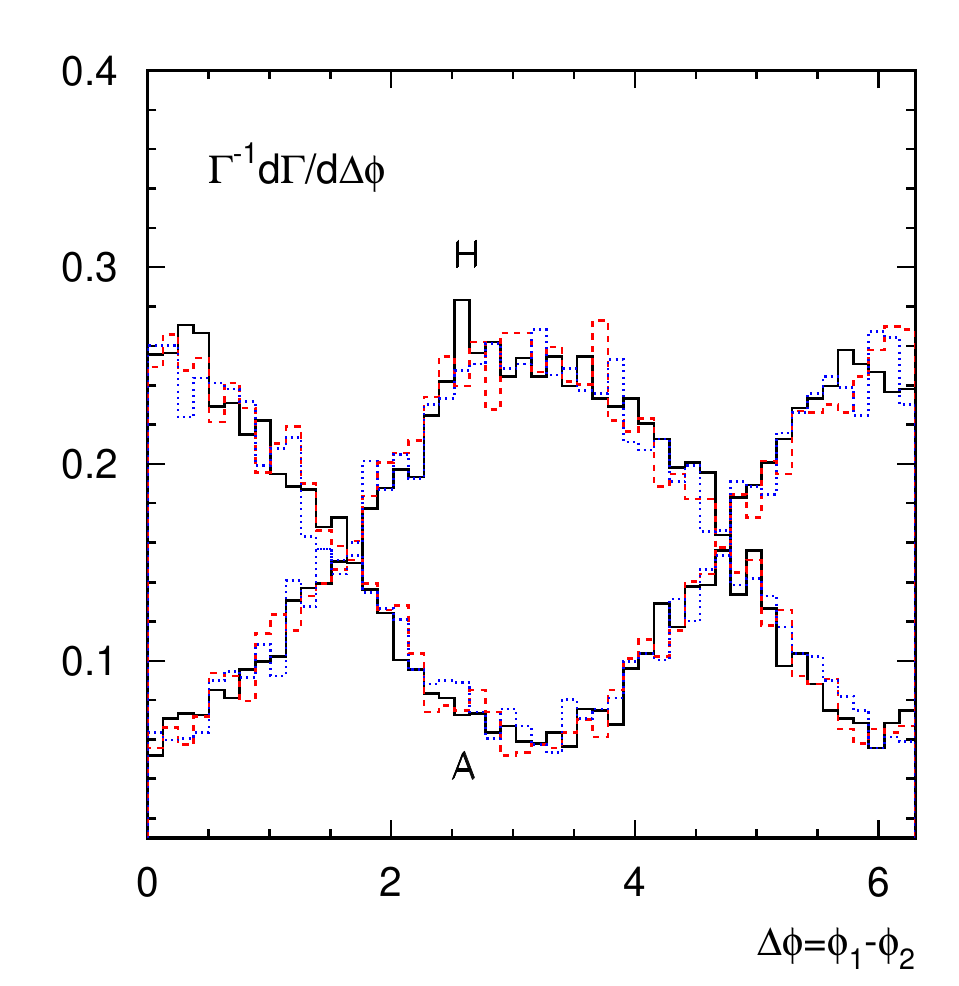}
 \caption{Azimuthal angle $\Delta\phi$ correlation in
 $pp\to H/A\to\tau^-(\to\pi^-\nu)\,\tau^+(\to\pi^+\bar\nu)$ by three
 different methods; A.~propagator (solid), B.~without density matrix
 (dashed), and C.~with density matrix (dotted).}
\label{fig:dphipphim123}
\end{figure}

Figure~\ref{fig:dphipphim123} shows the azimuthal angle $\Delta\phi$
correlation in 
$pp\to H/A\to\tau^-(\to\pi^-\nu)\,\tau^+(\to\pi^+\bar\nu)$ for the
comparison of the three methods, and the same results are reproduced
as expected.

\section{Summary}\label{sec:summary}

In this paper, we have implemented the main hadronic tau decays,
$\tau\to\nu_{\tau}+\pi$, $2\pi$, and $3\pi$, by using the effective
vertices in the {\sc FeynRules} and {\sc MadGraph5} framework, and 
constructed {\sc TauDecay}, a library of helicity amplitudes to simulate
polarized tau decays. The model file allows us to simulate tau decays
when the on-shell tau production is kinematically forbidden, and the
stau decay in the stau-neutralino coannihilation region was discussed as
an application. We also demonstrated that all possible correlations
among the decay products of pair-produced taus through a $Z$-boson and a
scalar/pseudoscalar Higgs boson can be produced within our full-fledged
package. The program has been tested carefully by making use of the
standard tau decay library {\sc Tauola}.

\begin{acknowledgement}{\textit{Acknowledgements}}
We wish to thank C.~Duhr and P.~de Aquino for their help with
{\sc FeynRules}
and J.~Alwall, T.~Stelzer and H.~Yokoya for their help with
{\sc MadGraph5} and {\sc Tauola}.
K.M. also thank J.~Keaveney for his help with {\sc Root}.
We also thank J.~Ellis, F.~Luo and K.~Olive for pointing out a typo
 in~\eqref{eq103}.  
J.N. acknowledges the warm hospitality extended to him by
 VUB.
T.L. is supported by the ARC Centre of Excellence for Particle Physics
 at the Terascale. 
The work presented here has been in part supported by the Concerted
 Research action
``Supersymmetric Models and their Signatures at the Large Hadron
 Collider'',  
the Strategic Research Program ``High Energy Physics" and
the Research Council of the Vrije Universiteit Brussel,
in part by the Belgian Federal Science Policy
Office through the Interuniversity Attraction Pole P7/37 and IAP VI/11,
and
by the Grant-in-Aid for Scientific Research (No. 20340064) from the
 Japan Society for the Promotion of Science.
\end{acknowledgement}

\paragraph{Note added.} 
The {\sc TauDecay} package is now supported by 
{\sc MadGraph5\_aMC@NLO} (v.2.2.2 or later)~\cite{Alwall:2014hca},
by using the form-factor option with the Fortran way; see more details
in the {\sc MadGraph5\_aMC@NLO} wiki page.

One can use the {\sc TauDecay} package by issuing the following commands
in a {\sc MadGraph5\_aMC@NLO} shell: e.g. 
\begin{verbatim}
 > import model sm-lepton_masses
 > add model taudecay_UFO
 > generate ta- > vt e- ve~ / w+ w-
 > add process ta- > vt mu- vm~ / w+ w-
 > add process ta- > vt pi-
 > add process ta- > vt pi- pi0
 > add process ta- > vt pi- pi0 pi0
 > add process ta- > vt pi- pi- pi+ EFT=1
 > output
 > launch
\end{verbatim}
By using the 'add model' command, the {\sc TauDecay} package can be
attached to any models. In addition to the hadronic tau decays, the
leptonic mode was implemented by using the effective four-fermion 
interaction.


\end{document}